\documentclass{IEEEtaes}

\usepackage{color,array,amsthm}
\usepackage{graphicx}

\jvol{XX}
\jnum{XX}
\jmonth{XXXXX}
\paper{1234567}
\pubyear{2026}
\doiinfo{TAES.2026.Doi Number}

\usepackage{color}
\usepackage[x11names]{xcolor}
\usepackage{mathtools}
\usepackage{amssymb}
\usepackage{lipsum}
\usepackage{float}
\usepackage{makecell}
\usepackage{longtable}
\usepackage{stfloats}
\usepackage{multirow}
\usepackage{amsmath}
\usepackage{bm}
\usepackage{algpseudocode}
\usepackage{booktabs}
\usepackage{balance}
\usepackage{soul}
\algdef{SE}[SUBALG]{Indent}{EndIndent}{}{\algorithmicend\ }%
\algtext*{Indent}
\algtext*{EndIndent}
\usepackage{threeparttable}
\allowdisplaybreaks
\usepackage{graphicx}
\usepackage{subfigure}
\usepackage[table]{xcolor}
\usepackage{hhline}
\usepackage[linesnumbered,ruled,vlined]{algorithm2e}

\usepackage{placeins}
\usepackage{esvect}
\usepackage{esint}

\begin{document}

\title{Multipath Extended Target Tracking with Labeled Random Finite Sets}

\author{Guanhua Ding}
\member{Student Member, IEEE}
\affil{Beihang University, Beijing, China} 

\author{Qinchen Wu}
\member{Student Member, IEEE}
\affil{Beihang University, Beijing, China} 

\author{Jinping Sun}
\member{Member, IEEE}
\affil{Beihang University, Beijing, China} 

\author{Yanping Wang}
\member{Member, IEEE}
\affil{North China University of Technology, Beijing, China} 

\author{Bing Zhu}
\member{Senior Member, IEEE}
\affil{Beihang University, Beijing, China} 

\author{Guoqiang Mao}
\member{Fellow, IEEE}
\affil{Southeast University, Nanjing, China}

\receiveddate{
This work has been submitted to the IEEE for possible publication. Copyright may be transferred without notice, after which this version may no longer be accessible. 
The work of Guanhua Ding, Qinchen Wu, and Jinping Sun was supported by the National Natural Science Foundation of China under Grants 62131001 and 62171029. The work of Bing Zhu was supported by the National Natural Science Foundation of China, Grant 62573019.}

\authoraddress{Authors' addresses: Guanhua Ding, Qinchen Wu, and Jinping Sun are with the School of Electronic Information Engineering, Beihang University, Beijing 100191, China, E-mail: (\url{buaadgh@buaa.edu.cn}, \url{wuqinchen@buaa.edu.cn}, \url{sunjinping@buaa.edu.cn}); Yanping Wang is with the School of Artificial Intelligence and Computer Science, North China University of Technology, Beijing 100144, China, E-mail: (\url{wangyp@ncut.edu.cn}). Bing Zhu is with the School of Automation Science and Electrical Engineering, Beihang University, Beijing 100191, China, E-mail: (\url{zhubing@buaa.edu.cn}). Guoqiang Mao is with Research Laboratory of Smart Driving and Intelligent Transportation Systems, Southeast University, Nanjing 210096, China, E-mail: (\url{g.mao@ieee.org}). \textit{(Corresponding author: Jinping Sun.)}}


\markboth{DING ET AL.}{MULTIPATH EXTENDED TARGET TRACKING WITH LABELED RANDOM FINITE SETS}
\maketitle

\begin{abstract}High-resolution radar sensors are critical for autonomous systems but pose significant challenges to traditional tracking algorithms due to the generation of multiple measurements per object and the presence of multipath effects.
Existing solutions often rely on the point target assumption or treat multipath measurements as clutter, whereas current extended target trackers often lack the capability to maintain trajectory continuity in complex multipath environments.
To address these limitations, this paper proposes the multipath extended target generalized labeled multi-Bernoulli (MPET-GLMB) filter.
A unified Bayesian framework based on labeled random finite set theory is derived to jointly model target existence, measurement partitioning, and the association between measurements, targets, and propagation paths.
This formulation enables simultaneous trajectory estimation for both targets and reflectors without requiring heuristic post-processing.
To enhance computational efficiency, a joint prediction and update implementation based on Gibbs sampling is developed.
Furthermore, a measurement-driven adaptive birth model is introduced to initialize tracks without prior knowledge of target positions.
Experimental results from simulated scenarios and real-world automotive radar data demonstrate that the proposed filter outperforms state-of-the-art methods, achieving superior state estimation accuracy and robust trajectory maintenance in dynamic multipath environments.
\end{abstract}


\section{INTRODUCTION}
M{\scshape ultiple} target tracking (MTT) is a fundamental task in autonomous driving, intelligent transportation, and robotics.
Traditional MTT formulations rely on the \textit{point target} assumption, which states that each target generates at most one measurement per sensor scan.
This assumption underlies classical MTT algorithms, including global nearest neighbor (GNN) \cite{bar2011tracking}, joint probabilistic data association (JPDA) \cite{fortmannJPDA1983}, multiple hypothesis tracking (MHT) \cite{reidMHT1979}, and random finite set (RFS)-based filters \cite{papiGeneralizedLabeledMultiBernoulli2015,liuGNNPMBSimpleEffective2022,dingOptiPMB2025}.
With the advent of high-resolution sensors such as LiDAR and mmWave radar, the point-target assumption has become largely invalid.
When the target physical dimensions exceed the sensor's resolution, reflections arise from multiple scattering centers, producing clusters of measurements rather than single points.
This necessitates joint estimation of target centroid kinematics and spatial extent (size and orientation), motivating the development of multiple \textit{extended target} tracking (METT) methods.
Various extended target models, including the random matrix \cite{feldmannTrackingExtendedObjects2011,meyerScalableDetectionTracking2021,xiaTrajectoryPMBFilters2023,liu4DEOT2024}, random hypersurface \cite{baumRHM2009}, Gaussian process \cite{baerveldtPMBMGaussianProcess2023}, and probabilistic measurement-region association \cite{dingFusion2024}, have been explored for radar- and LiDAR-based METT.

When radar sensors operate in geometrically complex environments, such as urban streets and indoor warehouses, the free-space line-of-sight (LOS) propagation assumption is frequently violated.
Electromagnetic waves may reflect off static structures, such as buildings and fences, before returning to the receiver, resulting in multipath propagation that creates \textit{ghost targets} at virtual locations. These effects pose significant challenges for data association and track management.

A straightforward approach to mitigate multipath effects is to suppress multipath return signals through antenna design and signal processing \cite{visentinMultipathDoaPolar2017,zhengDetectionGhostTargets2024} or discard ghost measurements using gating \cite{bar2011tracking} or machine learning-based identification methods \cite{prophetInstantaneousGhostDetection2019,chamseddineGhostTargetDetection2021}.
However, these approaches discard valuable information contained in multipath measurements.
In scenarios where the direct LOS path is occluded, multipath reflections may provide the only evidence of a target’s presence \cite{scheinerSeeingStreetCorners2020}.
Consequently, recent studies have sought to exploit multipath measurements as non-line-of-sight (NLOS) information rather than discarding them.
Several multi-detection (MD) tracking methods based on conventional JPDA and MHT filters have been developed for over-the-horizon radar (OTHR) and underwater sonar multipath propagation models \cite{liuUnderwaterTargetTracking2020,habtemariamMultiDetectionJPDA2013,sathyanMHTmultipath2013}.
Yang et al. proposed the MP-GLMB filter based on the generalized labeled multi-Bernoulli (GLMB) RFS for OTHR multipath tracking \cite{yangMultipathGeneralizedLabeled2018a}, followed by a more efficient Gibbs sampling-based implementation in \cite{wangRadarGhostTarget2021}.
These methods, however, rely on accurate prior knowledge of the multipath propagation model, which is often unavailable in dynamic traffic and indoor sensing environments.
A joint target-tracking and wall-estimation method was proposed in \cite{fengMultipathGhostRecognition2024}, where multipath measurements are identified using a Hough transform-based ghost recognition algorithm and unknown wall parameters are estimated via an extended Kalman filter (EKF).
Nevertheless, all the aforementioned methods assume point targets and are therefore unsuitable for high-resolution radar sensors.

For extended targets tracking with multipath measurements, the MP-RM-PDA filter \cite{liuExtendedTargetTracking2021} adopts random matrix modeling and explicitly enumerates possible multipath association events within a probabilistic data association framework, resulting in high computational complexity and requiring well-separated targets with known cardinality a priori.
The MP-ET-RFS tracker \cite{shenRFSbasedExtendedTarget2017} employs a two-stage strategy, using a probability hypothesis density (PHD) filter to pre-process measurements, followed by a multipath Bernoulli filter for final estimation.
Although this design reduces computational cost, the lack of explicit extent estimation can degrade tracking performance.
The MP-ET-PHD filter \cite{liuRFSBasedMultipleExtended2023} provides a state-of-the-art RFS-based solution that jointly estimates the kinematic state and spatial extent of an unknown number of targets using multipath measurements.
However, as a PHD-based method, it produces only unlabeled state estimates, preventing direct trajectory extraction without heuristic post-processing and often leading to unstable cardinality estimates \cite{lundquistGGIWCPHD2013}.
Moreover, these RFS-based methods generally rely on precise prior knowledge of multipath propagation models, which limits their applicability in complex and dynamic environments.

This study proposes the multipath extended target generalized labeled multi-Bernoulli (MPET-GLMB) filter, a unified Bayesian framework designed to address the aforementioned limitations.
To the best of our knowledge, this is the first work to effectively integrate a labeled RFS formulation with extended target models in dynamic multipath environments.
The main contributions of this work are summarized as follows:
\begin{itemize}
    \item A GLMB recursion is derived that jointly accounts for uncertainties in target existence, measurement partitioning, and measurement–target–path association. It enables simultaneous trajectory estimation of both targets and reflectors, providing a closed-form solution to multiple extended target tracking in multipath environments.

    \item A Gibbs sampling-based joint prediction and update implementation of the MPET-GLMB filter is developed, eliminating inefficient truncation procedures and resulting in a computationally tractable solution.
    
    \item A measurement-driven adaptive birth model is proposed for robust trajectory initialization that identifies potential reflectors and automatically determines the parameters of newborn targets.
    
    \item A comprehensive evaluation is conducted, in which the MPET-GLMB filter is benchmarked against the state-of-the-art MP-ET-PHD filter using both simulated and real-world radar data. The results demonstrate the superior state estimation accuracy and trajectory continuity of the proposed method in dynamic multipath environments.
    
\end{itemize}

The remainder of this paper is organized as follows.
Section~\ref{II.Background} provides background on the radar multipath propagation model and labeled RFS-based tracking.
Section~\ref{Section III} presents the mathematical derivation of the proposed MPET-GLMB filter, and Section~\ref{Section IV} describes its efficient implementation.
Experimental results and analysis are reported in Section~\ref{Section V}.
Finally, conclusions are drawn in Section~\ref{Section VI}.

\section{Background}
\label{II.Background}

\subsection{Radar Multipath Propagation Model}
\label{Radar Multipath Propagation Model}

\begin{figure}[t]
    \centering
    \includegraphics[width=0.8\linewidth]{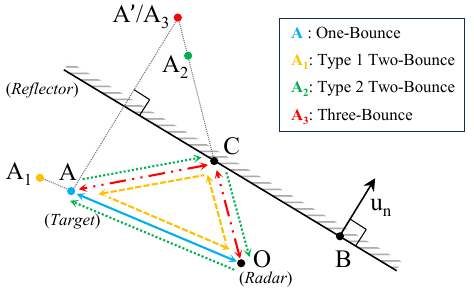}
    \caption{Radar multipath propagation geometry.}
    \label{fig1}
\end{figure}

According to standard extended target measurement models, each target measurement originates from a scattering center distributed on the target's spatial extent \cite{feldmannTrackingExtendedObjects2011,beardMultipleExtendedTarget2016a,lanTrackingExtendedObject}. Therefore, defining a multipath propagation model for point targets provides the basis for solving the METT problem. 

Consider the geometry illustrated in Fig.~\ref{fig1}, where a radar located at $\mathrm{O}=[o_\mathrm{x},o_\mathrm{y}]^\mathrm{T}$ observes a target at $\mathrm{A}=[a_\mathrm{x},a_\mathrm{y}]^\mathrm{T}$ with respective velocities $v^\mathrm{O}=[v_\mathrm{x}^\mathrm{O},v_\mathrm{y}^\mathrm{O}]^\mathrm{T}$ and $v^\mathrm{A}=[v_\mathrm{x}^\mathrm{A},v_\mathrm{y}^\mathrm{A}]^\mathrm{T}$.
The signal is reflected at a specular point $\mathrm{C}=[c_\mathrm{x},c_\mathrm{y}]^\mathrm{T}$ on a stationary reflective surface $\mathfrak{S}$, resulting in multipath effects.
Without loss of generality, the reflective surface $\mathfrak{S}$ is modeled as a line characterized by a point $\mathrm{B}=[b_\mathrm{x},b_\mathrm{y}]^\mathrm{T}$ and a unit normal vector $u_\mathrm{n}=[u_\mathrm{x},u_\mathrm{y}]^\mathrm{T}$.
As shown in Fig.~\ref{fig1}, the mirrored target $\mathrm{A}'$ and its velocity $v^{\mathrm{A}'}$ are given by
\begin{equation}
    \begin{gathered}
        \mathrm{A}' = \mathrm{A} - \vv{\mathrm{A'A}} = \mathrm{A} - 2(\vv{\mathrm{B}\mathrm{A}}\cdot u_\mathrm{n})u_\mathrm{n}, \\
        v^{\mathrm{A}'} = {\mathrm{d}\mathrm{A}'}/{\mathrm{dt}} = v^\mathrm{A} - 2(v^\mathrm{A} \cdot u_\mathrm{n})u_\mathrm{n},
    \end{gathered} \notag
\end{equation}
where $p\cdot q$ denotes the dot product, $\vv{pq}\triangleq q-p$, and $\mathrm{d}(\cdot)/\mathrm{dt}$ denotes the time derivative.

The specular point $\mathrm{C}$ is the intersection of the ray $\vv{\mathrm{OA}'}$ and the reflective surface $\mathfrak{S}$.
Let $u_\mathrm{t}=[-u_\mathrm{y},u_\mathrm{x}]^\mathrm{T}$ be the unit tangent vector of $\mathfrak{S}$.
The point $\mathrm{C}$ is obtained by solving $\mathrm{B}+s_1u_\mathrm{t} = \mathrm{O}+s_2\vv{\mathrm{OA}'}$ for scalars $s_1$ and $s_2$.
Eliminating $s_2$ yields
\begin{equation}
    s_1 = -(\vv{\mathrm{OA}'} \ast \vv{\mathrm{OB}})/(\vv{\mathrm{OA}'} \ast u_\mathrm{t}), \notag
\end{equation}
where $p\ast q \triangleq p_\mathrm{x}q_\mathrm{y}-p_\mathrm{y}q_\mathrm{x}$.
The specular point is then given by $\mathrm{C}=\mathrm{B}+s_1u_\mathrm{t}$.

With the above model, four propagation paths are identified:
\begin{itemize}
    \item Direct path (one-bounce) $\mathfrak{P}_0$: $\mathrm{O}\to \mathrm{A}\to \mathrm{O}$;
    \item Type~1 two-bounce path $\mathfrak{P}_1$: $\mathrm{O}\to \mathrm{C}\to \mathrm{A}\to \mathrm{O}$;
    \item Type~2 two-bounce path $\mathfrak{P}_2$: $\mathrm{O}\to \mathrm{A}\to \mathrm{C}\to \mathrm{O}$;
    \item Three-bounce path $\mathfrak{P}_3$: $\mathrm{O}\to \mathrm{C}\to \mathrm{A}\to \mathrm{C}\to \mathrm{O}$.
\end{itemize}
Higher-order multipath effects caused by multiple reflective surfaces are not considered in this study.

Assume that the radar measures range $r$, azimuth $\phi$, and Doppler velocity $\dot r$.
For the direct propagation path $\mathfrak{P}_0$, the ideal measurement $z_0=[r_0,\phi_0,\dot{r}_0]^\mathrm{T}$ is given by:
\begin{equation}
    z_0 = \left[ \|\vv{\mathrm{OA}}\|, \ \angle(\vv{\mathrm{OA}}), \ (v^\mathrm{A} - v^\mathrm{O})\cdot \left(\vv{\mathrm{OA}}/{\|\vv{\mathrm{OA}}\|}\right) \right]^\mathrm{T}, \notag
\end{equation}
where $\|\cdot\|$ denotes the Euclidean norm, and $\angle(\vv{\mathrm{OA}})$ returns the angle between $\vv{\mathrm{OA}}$ and the $\mathrm{x}$-axis.

The three-bounce propagation path $\mathfrak{P}_3$ is geometrically equivalent to a direct path toward the mirrored target $\mathrm{A}'$.
Accordingly, its ideal measurement $z_3$ is obtained by substituting $\mathrm{A}$ and $v^\mathrm{A}$ in $z_0$ with $\mathrm{A}'$ and $v^{\mathrm{A}'}$, respectively.
For the two-bounce paths $\mathfrak{P}_1$ and $\mathfrak{P}_2$, the range and Doppler are averaged over the $\mathfrak{P}_0$ and $\mathfrak{P}_3$ paths \cite{fengMultipathGhostRecognition2024}.
Consequently, the path $\mathfrak{P}_1$ generates a ghost target $\mathrm{A}_1$ sharing the azimuth of $\mathrm{A}$, while $\mathfrak{P}_2$ creates a ghost target $\mathrm{A}_2$ sharing the azimuth of $\mathrm{A}'$.
The ideal measurements are given by
\begin{equation}
    \begin{aligned}
        z_1 = \left[ \frac{r_0+r_3}{2}, \ \phi_0, \ \frac{\dot{r}_0+\dot{r}_3}{2} \right]^\mathrm{T},\ 
        z_2 = \left[ r_1, \ \phi_3, \ \dot{r}_1 \right]^\mathrm{T}. \notag
    \end{aligned}
\end{equation}

\subsection{Multi-Target Tracking with Labeled RFS}

RFS-based multi-target Bayes filtering provides a flexible and rigorous framework for estimating the states of multiple targets from noisy measurements \cite{voOverviewMultiObjectEstimation2024}.
Rather than modeling each target's state individually, the multi-target state is represented as a finite set, enabling principled handling of missed detections, false alarms (clutter), and unknown data association.
Labeled RFSs \cite{voLabeledRandomFinite2013} are widely adopted in MTT, as the labels allow online trajectory estimation without heuristic post-processing.
Following conventional notation, unlabeled vectors and scalars are denoted by lower case letters (e.g., $x$). Unlabeled sets and matrices are in upper case (e.g., $X$). Labeled states and densities are in boldface (e.g., $\mathbf{x}$, $\mathbf{X}$, and $\boldsymbol{\pi}$).
The multi-target exponential of a function is given by $h^{X}\triangleq\prod_{x\in X} h(x)$, with $h^{\emptyset}=1$. The inner product is denoted by $\langle f,g\rangle \triangleq \int f(x) g(x) \mathrm{d}x$.
The generalized Kronecker delta and the inclusion function are
\begin{align}
    \delta_Y(X) = \begin{cases} 1 & \text{if }X=Y\\ 0 & \text{otherwise} \end{cases}, \quad
    \mathit{1}_Y(X) &= \begin{cases} 1 & \text{if } X\subseteq Y\\ 0 & \text{otherwise} \end{cases}. \notag
\end{align}

\textit{Definition 1}: A labeled RFS $\mathbf{X}$ on the state space $\mathbb{X}$ and label space $\mathbb{L}$ consists of elements with distinct labels \cite{voLabeledRandomFinite2013}.
With the label projection $\mathcal{L}(x,\ell)=\ell$, the distinct label indicator is defined by $\Delta(\mathbf{X})=\delta_{|\mathbf{X}|}(|\mathcal{L}(\mathbf{X})|)$, where $|\mathbf{X}|$ denotes set cardinality and $\mathcal{L}(\mathbf{X})$ is the set of labels.

At time $k$, the multi-object state $\mathbf{X}_k$ is partially observed through a set of measurements $Z_k\subset\mathbb{Z}$. The measurement history is denoted as $Z_{1:k-1}=(Z_1,\ldots,Z_{k-1})$.
Given the prior density $\boldsymbol{\pi}_{k-1}(\mathbf{X}_{k-1}\mid Z_{1:k-1})$, the prediction density is computed via the Chapman--Kolmogorov equation \cite{voLabeledRandomFinite2013}
\begin{align}
\label{Chapman Kolmogorov}
    \boldsymbol{\pi}&_{k|k-1}(\mathbf{X}_k|Z_{1:k-1}) \\
    &= \int f_{k|k-1}(\mathbf{X}_k|\mathbf{X}_{k-1}) \boldsymbol{\pi}_{k-1}(\mathbf{X}_{k-1}|Z_{1:k-1}) \delta \mathbf{X}_{k-1},\notag
\end{align}
where $f_{k|k-1}(\mathbf{X}_k\mid\mathbf{X}_{k-1})$ is the multi-target transition kernel.
The posterior density is obtained using Bayes' rule
\begin{equation}
\label{Bayes equation}
    \boldsymbol{\pi}_{k}(\mathbf{X}_k|Z_{1:k})=\frac{\boldsymbol{\pi}_{k|k-1}(\mathbf{X}_k|Z_{1:k-1}) g_k(Z_k|\mathbf{X}_k)}{\int \boldsymbol{\pi}_{k|k-1}(\mathbf{X}_k|Z_{1:k-1}) g_k(Z_k|\mathbf{X}_k) \delta \mathbf{X}},
\end{equation}
where $g_k(Z_k|\mathbf{X}_k)$ denotes the multi-target likelihood.
Labeled target states can be extracted directly from the resulting posterior density.
For notational simplicity, the time index $k$ and the conditioning on past measurements are omitted hereafter, and the subscript $k|k-1$ is abbreviated as $+$.

A conjugate prior is required in multi-target Bayes filtering to obtain closed-form solutions to \eqref{Chapman Kolmogorov} and \eqref{Bayes equation}.
The GLMB density is a conjugate prior widely used in labeled RFS-based trackers \cite{beardMultipleExtendedTarget2016a,voRFSprior2011,deuschLabeledMultiBernoulliFilter2014,nguyenTrackingCellsTheir2021}.

\textit{Definition 2}: A GLMB RFS $\mathbf{X}$ on $\mathbb{X}\times\mathbb{L}$ has density
\begin{align}
    \boldsymbol{\pi}(\mathbf{X})= \Delta(\mathbf{X})\sum_{c\in \mathbb{C}} w^{(c)}(\mathcal{L}(\mathbf{X})) [p^{(c)}]^\mathbf{X},\notag
\end{align}
where $\mathbb{C}$ is a discrete index set \cite{beardMultipleExtendedTarget2016a}.
The weights satisfy $\sum_{J\subseteq \mathbb{L}} \sum_{c\in\mathbb{C}} w^{(c)}(J)=1$, and each single-target densities satisfy $\int_{x\in\mathbb{X}} p^{(c)}(x,\ell)\,\mathrm{d}x = 1$.

\textit{Definition 3}: A $\delta$-GLMB RFS $\mathbf{X}$ on $\mathbb{X}\times\mathbb{L}$ is a special case of the GLMB with
\begin{equation}
\begin{gathered}
    \mathbb{C} = \mathcal{F}(\mathbb{L})\times \Xi, \quad p^{(c)} = p^{(I,\xi)} = p^{(\xi)}, \\
    w^{(c)}(J) = w^{(I,\xi)}(J)= w^{(I,\xi)}\delta_I(J),
\end{gathered}\notag
\end{equation}
where $\mathcal{F}(\mathbb{L})$ denotes the set of all finite subsets of $\mathbb{L}$, and $\Xi$ denotes a discrete space \cite{voLabeledRandomFinite2013}.
The resulting density is:
\begin{align}
\label{prior GLMB}
    \boldsymbol{\pi}(\mathbf{X})=
    \Delta(\mathbf{X})
    \sum_{I\in \mathcal{F}(\mathbb{L}),\ \xi\in\Xi}
    w^{(I,\xi)}\delta_I(\mathcal{L}(\mathbf{X})) [p^{(\xi)}]^\mathbf{X}.
\end{align}
Each hypothesis $(I,\xi)$ consists of a label set $I$ and an association history $\xi$. The weight $w^{(I,\xi)}$ denotes the probability of this hypothesis, and $p^{(\xi)}(x,\ell)$ is the single-target density.

Existing GLMB filters address point targets with known multipath propagation models \cite{yangMultipathGeneralizedLabeled2018a} but do not handle extended targets with dynamic multipath effects.
In following sections, we derive a $\delta$-GLMB-based Bayesian filtering recursion for METT with dynamic multipath observations, and present an efficient implementation of the proposed filter.

\section{METT with Dynamic Multipath Observations}
\label{Section III}

\subsection{Observation Model of Multiple Extended Targets}

Conditioned on the multi-target state $\mathbf{X}$ and the measurement set $Z$, the multi-target likelihood $g(Z|\mathbf{X})$ is derived under the following assumptions:
\begin{itemize}
    \item[\textbf{A1.}]
    A target with state $\mathbf{x}\in\mathbf{X}$ is detected via the direct path $\mathfrak{P}_0$ with probability $P_\mathrm{d}(\mathbf{x})$. Its detection probability via path $\mathfrak{P}_{m\in M=\{1,2,3\}}$ is $P_\mathrm{d,R}(\mathbf{x},\mathbf{x}^\ast,m)$, where $\mathbf{x}^\ast\in \mathbf{X}$ is the state of another target providing the reflective surface. The misdetection probabilities are $q_\mathrm{d}(\mathbf{x})=1-P_\mathrm{d}(\mathbf{x})$ and $q_\mathrm{d,R}(\mathbf{x},\mathbf{x}^\ast,m)=1-P_\mathrm{d,R}(\mathbf{x},\mathbf{x}^\ast,m)$, respectively.
    \item[\textbf{A2.}]
    If detected via $\mathfrak{P}_0$, the target generates measurements $W\subseteq Z$ with likelihood $\tilde{g}_\mathrm{D}(W|\mathbf{x})$.
    If detected via $\mathfrak{P}_m$, the likelihood is $\tilde{g}_\mathrm{R}(W|\mathbf{x},\mathbf{x}^\ast,m)$.
    \item[\textbf{A3.}]
    Clutter is modeled as an independent Poisson RFS with intensity $\kappa(\cdot)$.  
    For a clutter set $C$, the likelihood is $g_\mathrm{c}(C)=e^{-\int \kappa(z)\,\mathrm{d}z}\kappa^C$ \cite{voRFSprior2011}.
\end{itemize}

By enumerating all multi-bounce propagation paths, the set of multi-bounce target state tuples is
\begin{align}
    \mathbf{T}(\mathbf{X},M)
    =
    \{(\mathbf{x}_i,\mathbf{x}_{j},m): i,j\in\mathcal{L}(\mathbf{X}),\ i\neq j,\ m\in M\},
    \notag
\end{align}
where each tuple $\mathbf{t}\in \mathbf{T}(\mathbf{X},M)$ is uniquely labeled by $\rho=(i,j,m)$ with label projection $\mathfrak{L}(t,\rho)=\rho$.

Let $\mathcal{P}^{S}(Z)$ denote the set of all partitions that separate $Z$ into exactly $S$ non-empty subsets.
For a partition $\mathcal{U}=\{\mathcal{U}_1,\dots,\mathcal{U}_{S}\}\in\mathcal{P}^{S}(Z)$, let $\Theta^{\mathcal{U}}_{\mathbf{X},\mathbf{T}}$ denote the set of admissible association mappings $(\theta,\varphi)$, where $\theta:\mathcal{L}(\mathbf{X})\to \{0,\ldots,|\mathcal{U}|\}$ denotes the direct path association and $\varphi:\mathfrak{L}(\mathbf{T}(\mathbf{X},M))\to \{0,\ldots,|\mathcal{U}|\}$ is the multi-bounce path association.
Validity requires that all non-zero assignments are unique and mutually disjoint, i.e., $\theta(\ell)\neq\varphi(\rho)$ whenever $\theta(\ell)>0$ and $\varphi(\rho)>0$.

\textit{Proposition 1}: The multi-target measurement likelihood is:
\begin{align}
\label{g(Z|X)}
    g(Z|\mathbf{X}) =& g_\mathrm{c}(Z) \quad\sum_{\mathclap{\substack{S=1,\dots,\mathcal{S}(\mathbf{X},M)+1\\ \mathcal{U}\in \mathcal{P}^{S}(Z),\ (\theta,\varphi)\in\Theta^{\mathcal{U}}_{\mathbf{X},\mathbf{T}}}}}\quad
    [\Psi_{\mathrm{D},\mathcal{U}}^{(\theta)}]^\mathbf{X}[\Psi_{\mathrm{R},\mathcal{U}}^{(\varphi)}]^{\mathbf{T}(\mathbf{X},M)}.
\end{align}
Here, $\mathcal{S}(\mathbf{X},M)=|\mathbf{X}|+|\mathbf{X}|(|\mathbf{X}|-1)|M|$ is the total number of possible propagation paths.
The functions $\Psi_{\mathrm{D},\mathcal{U}}^{(\theta)}(\mathbf{x})\triangleq\psi_{\mathrm{D},\mathcal{U}}^{(\theta(\mathcal{L}(\mathbf{x})))}(\mathbf{x})$, $\Psi_{\mathrm{R},\mathcal{U}}^{(\varphi)}(\mathbf{t})\triangleq\psi_{\mathrm{R},\mathcal{U}}^{(\varphi(\mathfrak{L}(\mathbf{t})))}(\mathbf{t})$ correspond to direct path and multi-bounce path associations, respectively.
The component likelihoods are:
\begin{subequations}
    \begin{align}
        &\psi_{\mathrm{D},\mathcal{U}}^{(s)}(\mathbf{x}) = \begin{cases}
            \frac{P_\mathrm{d}(\mathbf{x})\tilde{g}_\mathrm{D}(\mathcal{U}_{s}|\mathbf{x})}{\kappa^{\mathcal{U}_{s}}} & \text{if }s\in\{1:|\mathcal{U}|\}\\
            q_\mathrm{d}(\mathbf{x}) &\text{if } s=0
        \end{cases},\\
\label{psi_R}
        &\psi_{\mathrm{R},\mathcal{U}}^{(s)}(\mathbf{t}) = \begin{cases}
            \frac{P_\mathrm{d,R}(\mathbf{t}) \tilde{g}_\mathrm{R}(\mathcal{U}_{s}|\mathbf{t})}{\kappa^{\mathcal{U}_{s}}} &\text{if } s\in\{1:|\mathcal{U}|\}\\
            q_\mathrm{d,R}(\mathbf{t}) &\text{if } s=0
        \end{cases}.
    \end{align}
\end{subequations}

\textit{Proof}: Separate the measurement set $Z$ into disjoint subsets 
\begin{equation}
\begin{aligned}
    \mathcal{U}^\ast(Z)=\{C,Y_{n},B_{\rho}:n\in\mathcal{L}(\mathbf{X}),\rho\in\mathfrak{L}(\mathbf{T}),\uplus_{\mathcal{U}^\ast}=Z\}, \notag
\end{aligned}
\end{equation}
where $C$ denotes the clutter, $Y_{n}$ is the direct-path measurement set generated by target $\mathbf{x}_n$, and
$B_{\rho=(i,j,m)}$ is the multi-bounce measurement set of target $\mathbf{x}_i$ via propagation path $\mathfrak{P}_{m}$, with target $\mathbf{x}_j$ acting as the reflector.
Notably, $\mathcal{U}^*(Z)$ is not a partition of $Z$, as it may contain empty subsets.
Under Assumptions~A1--A3, the multi-target likelihood is written as a sum over all realizations of $\mathcal{U}^\ast(Z)$:
\begin{equation}
\label{original g(Z|X)}
    g(Z|\mathbf{X}) = \sum_{\mathcal{U}^\ast(Z)} g_\mathrm{c}(C) \prod_{n,\rho} g_\mathrm{D}(Y_{n}|\mathbf{x}_n) g_\mathrm{R}(B_{\rho}|\mathbf{t}_\rho),
\end{equation}
where $n\in\mathcal{L}(\mathbf{X})$ and $\rho\in\mathfrak{L}(\mathbf{T})$, with
\begin{align}
    g_\mathrm{D}(Y|\mathbf{x}) &= \begin{cases}
        P_\mathrm{d}(\mathbf{x}))\tilde{g}_\mathrm{D}(Y|\mathbf{x}) & \text{if } Y\neq\emptyset\\
        q_\mathrm{d}(\mathbf{x}) & \text{if } Y=\emptyset
    \end{cases}, \notag\\
    g_\mathrm{R}(B|\mathbf{t}) &= \begin{cases}
        P_\mathrm{d,R}(\mathbf{t}) \tilde{g}_\mathrm{R}(B|\mathbf{t}) & \text{if } B\neq\emptyset\\
        q_\mathrm{d,R}(\mathbf{t}) & \text{if } B=\emptyset \notag
    \end{cases}.
\end{align}
Separating the empty and non-empty subsets of $\mathcal{U}^\ast(Z)$ yields 
\begin{align}
    g(&Z|\mathbf{X})
    =\ \sum_{\mathclap{C\uplus W=Z}}\ g_\mathrm{c}(C) [q_\mathrm{d}]^{\mathbf{X}} [q_\mathrm{d,R}]^{\mathbf{T}(\mathbf{X},M)}
    \qquad \sum_{\mathclap{\substack{S=1,\dots,\mathcal{S}(\mathbf{X},M)\\ 
    \mathcal{U}\in \mathcal{P}^{S}(W), \
    (\theta,\varphi)\in\Theta^{\mathcal{U}}_{\mathbf{X},\mathbf{T}}}}} \notag\\
    &\times \prod_{\mathclap{\theta(n)>0}}\frac{P_\mathrm{d}(\mathbf{x}_n)\tilde{g}_\mathrm{D}(\mathcal{U}_{\theta(n)}|\mathbf{x}_n)}{q_\mathrm{d}(\mathbf{x}_n)} 
    \prod_{\mathclap{\varphi(\rho)>0}}\frac{P_\mathrm{d,R}(\mathbf{t}_\rho) \tilde{g}_\mathrm{R}(\mathcal{U}_{\varphi(\rho)}|\mathbf{t}_\rho)}{q_\mathrm{d,R}(\mathbf{t}_\rho) }.\notag
\end{align}
Since $C=Z-W$, the clutter can be absorbed into the measurement partition, resulting in
\begin{align}
    g(&Z|\mathbf{X})
    =g_\mathrm{c}(Z) [q_\mathrm{d}]^{\mathbf{X}} [q_\mathrm{d,R}]^{\mathbf{T}(\mathbf{X},M)}
    \qquad \sum_{\mathclap{\substack{S=1,\dots,\mathcal{S}(\mathbf{X},M)+1\\ 
    \mathcal{U}\in \mathcal{P}^{S}(Z),\ (\theta,\varphi)\in\Theta^{\mathcal{U}}_{\mathbf{X},\mathbf{T}}}}} \notag\\
    &\times \ \prod_{\mathclap{\theta(n)>0}}\frac{P_\mathrm{d}(\mathbf{x}_n)\tilde{g}_\mathrm{D}(\mathcal{U}_{\theta(n)}|\mathbf{x}_n)}{q_\mathrm{d}(\mathbf{x}_n) \kappa^{\mathcal{U}_{\theta(n)}}} \prod_{\mathclap{\varphi(\rho)>0}}\frac{P_\mathrm{d,R}(\mathbf{t}_\rho) \tilde{g}_\mathrm{R}(\mathcal{U}_{\varphi(\rho)}|\mathbf{t}_\rho)}{q_\mathrm{d,R}(\mathbf{t}_\rho) \kappa^{\mathcal{U}_{\varphi(\rho)}}}.\notag
\end{align}
Finally, by multiplying $[q_\mathrm{D}^\mathrm{d}]^{\mathbf{X}}$ and $[q_\mathrm{R}^\mathrm{d}]^{\mathbf{T}(\mathbf{X},M)}$ into the summation, the misdetection terms in the denominators cancel out, yielding the likelihood expression in~\eqref{g(Z|X)}.

\subsection{$\delta$-GLMB Filtering Recursion}
\label{GLMB Filtering Recursion}

Consider a $\delta$-GLMB multi-target prior density of the form \eqref{prior GLMB} and a newborn target state $\mathbf{X}_\mathrm{B}$ with density \cite{voLabeledRandomFinite2013}
\begin{align}
\label{birth LMB}
   \Delta(\mathbf{X}_\mathrm{B}) [\mathit{1}_{\mathbb{B_+}} r_{\mathrm{B}+}]^{\mathcal{L}(\mathbf{X}_\mathrm{B})} [1-r_{\mathrm{B}+}]^{\mathbb{B}_+ - \mathcal{L}(\mathbf{X}_\mathrm{B})}  [p_{\mathrm{B}+}]^{\mathbf{X}_\mathrm{B}},
\end{align}
where $\mathbb{B}_+$ is the label space of newborn targets, $r_{\mathrm{B}+}(\ell)$ is the birth probability of a target labeled by $\ell$, and $p_{\mathrm{B}+}(x,\ell)$ is the corresponding single-target state density.

Conditioned on $\mathbf{X}$, each target $\mathbf{x}\in\mathbf{X}$ survives independently with probability $P_\mathrm{S}(\mathbf{x})$ and transitions according to $f_+(\mathbf{x_+|\mathbf{x}})$. 
Under these assumptions, the $\delta$-GLMB prediction density is given by
\cite{voEfficientImplementationGeneralized2017}:
\begin{align}
\label{prediction density}
    \boldsymbol{\pi}_+(\mathbf{X}) = \Delta(\mathbf{X})\quad\sum_{\mathclap{\substack{\xi\in\Xi,\ I_+\in\mathcal{F}(\mathbb{L}\uplus\mathbb{B_+})}}}\quad w_+^{(I_+,\xi)} \delta_{I_+}(\mathcal{L}(\mathbf{X})) [\bar{p}_+^{(\xi)}]^\mathbf{X},
\end{align}
where
\begin{subequations}
\begin{align}
    \label{w_+}&w_+^{(I_+,\xi)} = [\mathit{1}_{\mathbb{B_+}} r_{\mathrm{B}+}]^{I_+\cap \mathbb{B}_+} [1-r_{\mathrm{B}+}]^{\mathbb{B}_+ - I_+}\sum_{I\in\mathcal{F}(\mathbb{L})}\\
    &\qquad\times \mathit{1}_{\mathcal{F}(I)}(I_+\cap I) [\bar{P}_\mathrm{S}^{(\xi)}]^{I_+\cap I} [1-\bar{P}_\mathrm{S}^{(\xi)}]^{I-I_+} w^{(I,\xi)}, \notag\\
    &\bar{p}_+^{(\xi)}(\mathbf{x}_+) = \mathit{1}_{\mathbb{B_+}}(\ell_+) p_{\mathrm{B}+}(\mathbf{x}_+) + \mathit{1}_{\mathbb{L}}(\ell_+) \\
        &\qquad\times {\langle p^{(\xi)}(\cdot,\ell_+), P_\mathrm{S}(\cdot,\ell_+) f_+(x_+|\cdot,\ell_+)\rangle}/ {\bar{P}_\mathrm{S}^{(\xi)}(\ell_+)}, \notag\\
    &\bar{P}_\mathrm{S}^{(\xi)}(\ell_+) = \langle p^{(\xi)}(\cdot,\ell_+), P_\mathrm{S}(\cdot,\ell_+) \rangle.
\end{align}
\end{subequations}

Multiplying \eqref{prediction density} by the likelihood \eqref{g(Z|X)} yields
\begin{align}
    \boldsymbol{\pi}_+(\mathbf{X}) g(Z|\mathbf{X}&) = g_\mathrm{c}(Z) 
    \Delta(\mathbf{X}) \sum_{\xi,I_+,S,\mathcal{U},\theta,\varphi}
    w_+^{(I_+,\xi)}  \notag\\
    &\times \delta_{I_+}(\mathcal{L}(\mathbf{X}))[\bar{p}_+^{(\xi)} \Psi_{\mathrm{D},\mathcal{U}}^{(\theta)}]^\mathbf{X} [\Psi_{\mathrm{R},\mathcal{U}}^{(\varphi)}]^{\mathbf{T}(\mathbf{X},M)}, \notag
\end{align}
where $(\xi,I_+)\in \Xi\times\mathcal{F}(\mathbb{L}\uplus\mathbb{B_+})$, $S=1,...,\mathcal{S}(\mathbf{X},M) + 1$, $\mathcal{U}\in \mathcal{P}^{S}(Z)$, $(\theta,\varphi)\in\Theta^{\mathcal{U}}_{\mathbf{X},\mathbf{T}}$.
Since $[\Psi_{\mathrm{R},\mathcal{U}}^{(\varphi)}]^{\mathbf{T}(\mathbf{X},M)}$ is not a multi-target exponential for $\mathbf{X}$, the posterior density $\boldsymbol{\pi}(\mathbf{X}|Z)\propto \boldsymbol{\pi}_+(\mathbf{X})g(Z|\mathbf{X})$ does not belong to the $\delta$-GLMB family, and approximations are thus required to restore conjugacy.

In traffic environments, reflective surfaces, such as walls and guardrails, exhibit distinct shapes, sizes, and motion characteristics that differ fundamentally from those of pedestrians and vehicles. 
Accordingly, separate state models are adopted for objects and reflectors, and the multi-target state is partitioned as $\mathbf{X} = \mathbf{X}^\mathrm{o} \uplus \mathbf{X}^\mathrm{r}$, where $\mathbf{x}^\mathrm{o} \in \mathbf{X}^\mathrm{o}$ denotes the state of an object and $\mathbf{x}^\mathrm{r} \in \mathbf{X}^\mathrm{r}$ is the state of a reflector. 
By neglecting multipath interactions among objects, the set of valid multi-bounce tuples reduces to $\tilde{\mathbf{T}}(\mathbf{X},M)=\{(\mathbf{x}_i,\mathbf{x}^\mathrm{r}_j,m):i\in \mathcal{L}(\mathbf{X}),j\in \mathcal{L}(\mathbf{X}^\mathrm{r}),i\neq j,m\in M\}=\{\tilde{\mathbf{t}}_\rho\}$.
Under these assumptions, the multi-target likelihood becomes
\begin{align}
\label{grouped g(Z|X)}
    g(Z|\mathbf{X})
    = g_\mathrm{c}(Z)\sum_{S, \mathcal{U},\theta,\varphi}
    [\Psi_{\mathrm{D},\mathcal{U}}^{(\theta)}]^\mathbf{X}[\Psi_{\mathrm{R},\mathcal{U}}^{(\varphi)}]^{\tilde{\mathbf{T}}(\mathbf{X},M)},
\end{align}
where $S=1,...,\mathcal{S}^\ast(\mathbf{X},M) + 1$, $\mathcal{S}^\ast(\mathbf{X},M)=|\mathbf{X}|+|\mathbf{X}^\mathrm{r}|(|\mathbf{X}|-1)|M|$ is the number of admissible propagation paths, $\mathcal{U}\in \mathcal{P}^{S}(Z)$, and $(\theta,\varphi)\in\Theta^{\mathcal{U}}_{\mathbf{X},\tilde{\mathbf{T}}}$.

Furthermore, to obtain a $\delta$-GLMB posterior, the function
$\psi_{\mathrm{R},\mathcal{U}}^{(s)}(\mathbf{t})$ in \eqref{psi_R}, defined for $\mathbf{t}=(\mathbf{x}_i,\mathbf{x}_j,m)$, is approximated by a function depending only on $\mathbf{x}_i$:
\begin{gather}
\label{approximated psi}  
\begin{aligned}
    \hat{\psi}_{\mathrm{R},\mathcal{U}}^{(s,j,m)}(\mathbf{x}_i) = \begin{cases}
            \frac{\hat{P}_\mathrm{d,R}^{(j,m)}(\mathbf{x}_i) \hat{g}_\mathrm{R}^{(j,m)}(\mathcal{U}_{s}|\mathbf{x}_i)}{\kappa^{\mathcal{U}_{s}}} & \text{if } s\in\{1:|\mathcal{U}|\},\\
            \hat{q}_\mathrm{d,R}^{(j,m)}(\mathbf{x}_i) & \text{if } s=0,
        \end{cases}
\end{aligned}
\raisetag{13pt}
\end{gather}
where $\hat{g}_\mathrm{R}^{(j,m)}(\mathcal{U}_{s}|\mathbf{x}_i)\triangleq\tilde{g}_\mathrm{R}(\mathcal{U}_{s}|\mathbf{x}_i,\hat{\mathbf{x}}^\mathrm{r}_j,m)$, $\hat{P}_\mathrm{d,R}^{(j,m)}(\mathbf{x}_i)\triangleq P_\mathrm{d,R}(\mathbf{x}_i,\hat{\mathbf{x}}^\mathrm{r}_j,m)$, and $\hat{q}_\mathrm{d,R}^{(j,m)}(\mathbf{x}_i)=1-\hat{P}_\mathrm{d,R}^{(j,m)}(\mathbf{x}_i)$.
Here, the estimated reflector state set $\hat{\mathbf{X}}^\mathrm{r}$ is treated as a fixed parameter, removing the explicit dependence on $\mathbf{x}_j^\mathrm{r}$.
Substituting \eqref{approximated psi} into \eqref{grouped g(Z|X)}, the multi-target likelihood is approximated by
\begin{equation}
\label{approximated g(Z|X)}
    \begin{aligned}
        \hat{g}(Z|\mathbf{X})
        = & g_\mathrm{c}(Z) \qquad \sum_{{\mathclap{\substack{S=1,\dots,\mathcal{S}^\ast(\mathbf{X},M) + 1\\ 
        \mathcal{U}\in \mathcal{P}^{S}(Z),\ (\theta,\varphi)\in\Theta^{\mathcal{U}}_{\mathbf{X},\tilde{\mathbf{T}}}}}}} \qquad
        [\hat{\Psi}_{\mathcal{U}}^{(\theta,\varphi)}]^\mathbf{X},
    \end{aligned}
\end{equation}
where 
\begin{align}
\label{psi}
    \hat\Psi_{\mathcal{U}}^{(\theta,\varphi)}(\mathbf{x}) =
        \psi_{\mathrm{D},\mathcal{U}}^{(\theta(\ell))}(\mathbf{x})
        \quad \prod_{\mathclap{\substack{j\in\mathcal{L}(\hat{\mathbf{X}}^\mathrm{r})-\{\ell\},\ m\in M}}}\quad
        \hat\psi_{\mathrm{R},\mathcal{U}}^{(\varphi(\ell,j,m),j,m)}(\mathbf{x}).
\end{align}

\textit{Proposition 2}: Given the predicted $\delta$-GLMB density in \eqref{prediction density} and the approximated multi-target likelihood in \eqref{approximated g(Z|X)}, the posterior multi-target density is a $\delta$-GLMB of the form
\begin{equation}
\label{update density}
    \boldsymbol{\pi}(\mathbf{X}|Z) = \Delta(\mathbf{X})\ \ \sum_{\mathclap{\substack{\xi, I_+, S, \mathcal{U},\theta,\varphi}}}\ \ 
    w_{\mathcal{U}}^{(I_+,\xi,\theta,\varphi)} \delta_{I_+}(\mathcal{L}(\mathbf{X})) [p_{\mathcal{U}}^{(\xi,\theta,\varphi)}]^\mathbf{X},
\end{equation}
where $(\xi,I_+)\in\Xi\times\mathcal{F}(\mathbb{L}\uplus\mathbb{B_+})$, $S=1,\dots,\mathcal{S}^\ast(\mathbf{X},M) + 1$, $\mathcal{U}\in \mathcal{P}^{S}(Z)$, $(\theta,\varphi)\in\Theta^{\mathcal{U}}_{\mathbf{X},\tilde{\mathbf{T}}}$. 
The component weights and single-target posteriors are given by
\begin{subequations}
\label{GLMB update}
\begin{align}
     w_{\mathcal{U}}^{(I_+,\xi,\theta,\varphi)} &\propto w_+^{(I_+,\xi)} [\eta_{\mathcal{U}}^{(\xi,\theta,\varphi)}]^{I_+}\label{w_Z},\\
     p_{\mathcal{U}}^{(\xi,\theta,\varphi)}(x,\ell) &= {\bar{p}_+^{(\xi)}(x,\ell) \hat\Psi_{\mathcal{U}}^{(\theta,\varphi)}(x,\ell)}/{\eta_{\mathcal{U}}^{(\xi,\theta,\varphi)}(\ell)},\\
     \eta_{\mathcal{U}}^{(\xi,\theta,\varphi)}(\ell) &= \int \bar{p}_+^{(\xi)}(x,\ell) \hat\Psi_{\mathcal{U}}^{(\theta,\varphi)}(x,\ell) \mathrm{d}x.
\end{align}
\end{subequations}

\textit{Proof}: The product of the prediction density $\boldsymbol{\pi}_+(\mathbf{X})$ and the approximated likelihood $\hat{g}(Z|\mathbf{X})$ is
\begin{align}
    &\boldsymbol{\pi}_+(\mathbf{X}) \hat{g}(Z|\mathbf{X}) \notag\\
    &= g_\mathrm{c}(Z)\Delta(\mathbf{X})\sum_{\substack{\xi, I_+, S, \mathcal{U},\theta,\varphi}}  
    w_+^{(I_+,\xi)} \delta_{I_+}(\mathcal{L}(\mathbf{X})) \notag
    [\bar{p}_+^{(\xi)} \hat\Psi_{\mathcal{U}}^{(\theta,\varphi)}]^{\mathbf{X}}\notag\\
    &= g_\mathrm{c}(Z)\Delta(\mathbf{X})\sum_{\substack{\xi, I_+, S, \mathcal{U},\theta,\varphi}}  
    w_+^{(I_+,\xi)} \delta_{I_+}(\mathcal{L}(\mathbf{X}))
    [\eta_{\mathcal{U}}^{(\xi,\theta,\varphi)}]^{\mathcal{L}(\mathbf{X})}\notag\\
    &\quad \times [p_{\mathcal{U}}^{(\xi,\theta,\varphi)}(\cdot)]^{\mathbf{X}}\notag\\
    &= g_\mathrm{c}(Z)\Delta(\mathbf{X})\sum_{{\substack{\xi, I_+, S, \mathcal{U},\theta,\varphi}}}
    w_+^{(I_+,\xi)} [\eta_{\mathcal{U}}^{(\xi,\theta,\varphi)}]^{I_+} \delta_{I_+}(\mathcal{L}(\mathbf{X})) \notag\\
    &\quad \times[p_{\mathcal{U}}^{(\xi,\theta,\varphi)}]^{\mathbf{X}}.\notag
\end{align}
The normalizing factor is given by the set integral
\begin{align}
    \int &\boldsymbol{\pi}_+(\mathbf{X}) \hat{g}(Z|\mathbf{X}) \delta\mathbf{X}\notag\\
    &= g_\mathrm{c}(Z) \sum_{\xi, I_+} w_+^{(I_+,\xi)}
    \int \Delta(\mathbf{X})  \delta_{I_+}(\mathcal{L}(\mathbf{X}))
    \sum_{\substack{S,\mathcal{U},\theta,\varphi}}  \notag\\
    &\quad \times [\bar{p}_+^{(\xi)}(\cdot) \hat\Psi_{\mathcal{U}}^{(\theta,\varphi)}(\cdot)]^{\mathbf{X}} \delta\mathbf{X}\notag\\
    &= g_\mathrm{c}(Z) \sum_{\xi, I_+} w_+^{(I_+,\xi)}
    \sum_{L\subseteq \mathbb{L}\uplus\mathbb{B_+}} \delta_{I_+}(L)
    \sum_{\substack{S,\mathcal{U},\theta,\varphi}}  \notag\\
    &\quad \times \left[\int \bar{p}_+^{(\xi)}(x,\cdot)  \hat\Psi_{\mathcal{U}}^{(\theta,\varphi)}(x,\cdot) \mathrm{d}x \right]^L \notag\\
    &= g_\mathrm{c}(Z) \sum_{\xi, I_+} w_+^{(I_+,\xi)} 
    \sum_{\substack{S,\mathcal{U},\theta,\varphi}} [\eta_{\mathcal{U}}^{(\xi,\theta,\varphi)}]^{I_+}, \notag
\end{align}
where Lemma 3 in \cite{voLabeledRandomFinite2013} is applied. Substituting the above results into Bayes' rule \eqref{Bayes equation} yields the posterior \eqref{update density}.

\subsection{Extended Target Modeling}

In this study, the object states are modeled by the gamma Gaussian inverse Wishart (GGIW) density \cite{beardMultipleExtendedTarget2016a, lanExtendedObjectGroupTargetTracking2019}, while the reflectors are represented by a uniform stick model.

\subsubsection{GGIW Model}
\label{GGIW Model}

Let $W$ denote the set of measurements generated by an object with state $x=(x_\gamma,x_\mathrm{K},x_\mathrm{E})$. 
The number of measurements is assumed to follow a Poisson distribution $\mathcal{PS}(|W|;x_\gamma)$. Poisson rate $x_\gamma>0$ follows a gamma distribution with probability density function (PDF)
\begin{equation}
    \begin{aligned}
        \mathcal{G}(x_\gamma;\alpha, \beta)=\frac{\beta^\alpha}{\Gamma(\alpha)} x^{\alpha-1}_\gamma e^{-\beta x_\gamma}, \notag
    \end{aligned}
\end{equation}
where $\alpha >0$ is the shape parameter, and $\beta > 0$ is the rate parameter.
The kinematic state $x_\mathrm{K}=[\mathrm{p}_\mathrm{x},\mathrm{p}_\mathrm{y},v_\mathrm{x},v_\mathrm{y}]^\mathrm{T}\in\mathbb{R}^4$ is modeled as a Gaussian distribution $\mathcal{N}(x_\mathrm{K};\mu,P)$ with mean $\mu$ and covariance $P$. 
The spatial extent is represented by a symmetric positive definite matrix $x_\mathrm{E}\in\mathbb{S}^{2}$, which follows an inverse Wishart distribution
\begin{equation}
    \begin{aligned}
        \mathcal{IW}(x_\mathrm{E};\nu,V)=\frac{|V|^\frac{\nu}{2}|x_\mathrm{E}|^{\frac{-(\nu+3)}{2}}}{2^{\nu}\Gamma_{2}(\frac{\nu}{2})} e^{-\mathrm{tr}(Vx_\mathrm{E}^{-1})/2}, \notag
    \end{aligned}
\end{equation}
where $\nu>1$ is the degrees of freedom, $V\in\mathbb{S}^{2}$ is the scale matrix, $\Gamma_{2}(\cdot)$ is the multi-variate gamma function, and $\mathrm{tr}(\cdot)$ denotes the matrix trace. 
Combining the definitions above, the object state density is given by \cite{beardMultipleExtendedTarget2016a}:
\begin{equation}
\label{GGIW}
    \begin{aligned}
        p(x)=&\mathcal{G}(x_\gamma;\alpha, \beta) \mathcal{N}(x_\mathrm{K};\mu,P) \mathcal{IW}(x_\mathrm{E};\nu,V)\\
            =& \mathcal{GGIW}(x;\chi)
    \end{aligned}
\end{equation}
where $\chi=(\alpha,\beta,\mu,P,\nu,V)$ is the set of GGIW parameters. 

The kinematic state transition density is modeled as
\begin{align}
\label{kinematic predict}
    f_+(x_{\mathrm{K}+}|x_\mathrm{K})=\mathcal{N}(x_{\mathrm{K}+};F_+x_\mathrm{K},G_+ Q_+ G_+^\mathrm{T} ).
\end{align}
The near-constant-velocity motion model \cite{liSurveyManeuveringTargettracking2003} is adopted in this study, while other maneuver models can also be applied under the formulation of \eqref{kinematic predict}. The state transition matrix $F_+$, process noise covariance $Q_+$, and noise gain matrix $G_+$ are:
\begin{gather}
    F_{+}=\begin{bmatrix}
        1&\tau_+\\0&1
    \end{bmatrix}\otimes \mathrm{diag}(1,1),\quad Q_+=\mathrm{diag}(\sigma_\mathrm{x}^2,\sigma_\mathrm{y}^2),\notag\\ 
    G_+= [\tau_+^2/2, \tau_+]^\mathrm{T} \otimes \mathrm{diag}(1,1), \notag
\end{gather}
where $\otimes$ denotes the Kronecker product, $\tau_+$ is the time interval, and $\mathrm{diag}(\cdot)$ returns a diagonal matrix.
Following \cite{beardMultipleExtendedTarget2016a,lanTrackingExtendedObject,granstromNewPredictionExtended2014}, the predicted density $p_+(x)$ is approximated as $\mathcal{GGIW}(x;\chi_+)$ with parameters
\begin{gather}
    \alpha_+=\alpha/\epsilon,\ \beta_+=\beta/\epsilon,\notag\\ \mu_+=F_+\mu,\ P_+=F_+PF_+^\mathrm{T}+Q_+, \notag\\
    \nu_+=e^{-{\tau_+}/{\tau_0}}(\nu+3)-3,\ V_+=V{\nu_+}/{\nu}. \notag
\end{gather}
Here, $\epsilon$ is an exponential forgetting factor, where a larger $\epsilon$ increases the variance of $x_\gamma$. 
Similarly, $\tau_0$ is a temporal decay factor, where a larger $\tau_0$ results in a smaller variance in $x_\mathrm{E}$.

\subsubsection{Uniform Stick Model} 

The GGIW model, which assumes elliptical extents, is unsuitable for long stick-shaped reflectors. 
Therefore, a spatial distribution-based extent representation \cite{dingFusion2024,gilholmSpatialDistributionModel2005} is adopted, and reflectors are modeled as uniform sticks. 
Consider a reflector with state $x=(x_\gamma,x_\mathrm{K},x_\mathrm{E})$. 
The Poisson rate $x_\gamma$ and the kinematic state $x_\mathrm{K}$ are modeled identically to those of objects. 
The extent state is instead defined as $x_\mathrm{E}=[\mathfrak{h},\mathfrak{s},\mathfrak{e}]^\mathrm{T}\in\mathbb{R}^3$, where $\mathfrak{h}$ denotes the heading angle, and $(\mathfrak{s},\mathfrak{e})$ are the scalar offsets of the starting and ending points along the reflector.
Given the centroid position $\mathrm{C}_0=[\mathrm{p}_\mathrm{x},\mathrm{p}_\mathrm{y}]^\mathrm{T}$ and the extent state $x_\mathrm{E}$ of a reflector, the endpoints are computed as
\begin{align}
\label{c_start c_end}
    \mathrm{C}_\mathrm{start}=\mathrm{C}_0+\mathfrak{s}\begin{bmatrix}
        \cos(\mathfrak{h})\\ \sin(\mathfrak{h})
    \end{bmatrix},\ 
    \mathrm{C}_\mathrm{end}=\mathrm{C}_0+\mathfrak{e}\begin{bmatrix}
        \cos(\mathfrak{h})\\ \sin(\mathfrak{h})
    \end{bmatrix}.
\end{align}
The extent state transition density is modeled as
\begin{equation}
\label{particle transition density}
    f_+(x_{\mathrm{E}+}|x_\mathrm{E})= \mathcal{N}(x_{\mathrm{E}+};x_{\mathrm{E}},Q_{\mathrm{E}+}),
\end{equation}
where $Q_{\mathrm{E}+}=\mathrm{diag}(\sigma^2_\mathfrak{h}, \sigma^2_\mathfrak{s}, \sigma^2_\mathfrak{e})$ represents the process noise covariance matrix.

\subsection{Multipath Measurement Model}

Consider a set of measurements $W$ generated by an extended target with state $x=(x_\gamma,x_\mathrm{K},x_\mathrm{E})$. 
The likelihood function is $\tilde{g}(W|x)=\mathcal{PS}(|W|;x_\gamma)\prod_{z\in W}p(z|x)$ \cite{beardMultipleExtendedTarget2016a,granstromNewPredictionExtended2014}. 
Each measurement $z\in W$ originates from a reflective point with state $\check{x}=[\check{\mathrm{p}}_\mathrm{x}, \check{\mathrm{p}}_\mathrm{y}, \check{v}_\mathrm{x}.\, \check{v}_\mathrm{y}]^\mathrm{T}$.
Following Section~\hyperref[Radar Multipath Propagation Model]{II-A}, the measurement model under propagation path $\mathfrak{P}_m$ is
\begin{align}
    z=[r,\phi,\dot{r}]^\mathrm{T}=\begin{cases}
        h_m(\check{x},x_\mathrm{O}) + \omega & m=0,\\
        h_m(\check{x},x_\mathrm{O},x^\ast) + \omega & m\in\{1,2,3\},
    \end{cases} \notag
\end{align}
where $h_m(\cdot)$ is the nonlinear measurement function, $x_\mathrm{O}$ is the radar state, and $x^\ast$ denotes the reflector state. 
The measurement noise $\omega=[\omega_r,\omega_\phi,\omega_{\dot{r}}]^\mathrm{T}\sim\mathcal{N}(0,\tilde{R})$ has covariance matrix $\tilde{R}=\mathrm{diag}(\sigma^2_r,\sigma^2_\phi,\sigma^2_{\dot{r}})$. 
As discussed in Section~\hyperref[GLMB Filtering Recursion]{III-B}, the estimated reflector state $\hat x^\ast\in \hat{X}^\mathrm{r}$ is used to parameterize the reflective surface $\mathfrak{S}$.
The nonlinear measurement function is denoted by $\hat{h}_m(\cdot)$ for brevity.

Assuming reflective points are uniformly distributed over the target extent and share the centroid velocity, the single-measurement likelihood is
\begin{align}
\label{p_z_x}
    &p(z|x)=\int p(z|\check{x},x)p(\check{x}|x)\mathrm{d}\check{x} = \mathcal{N}(\dot{r};\hat{h}_m^{\dot{r}}(x_\mathrm{K}),\sigma^2_{\dot{r}}) \notag\\
    &\quad \times \int \mathcal{N}(z^{r\phi};\hat{h}_m^{r\phi}(\check{x}^\mathrm{p}),\mathrm{diag}(\sigma^2_r,\sigma^2_\phi))
    \mathcal{UN}(\check{x}^\mathrm{p};x) \mathrm{d}\check{x}^\mathrm{p},
\end{align}
where superscripts select vector components (e.g., $z^{r\phi}=[r,\phi]^\mathrm{T}$ denotes the range–azimuth measurement vector), and $\mathcal{UN}(\cdot)$ is the uniform density over the target extent.

\subsection{GGIW Update for Objects}
\label{GGIW Update for Objects}

For GGIW-modeled objects, the elliptical uniform distribution of reflective points is approximated by a Gaussian PDF $\mathcal{N}(\check{x}^\mathrm{p};x_\mathrm{K}^\mathrm{p}, \varrho x_\mathrm{E})$ with scaling parameter $\varrho=1/4$ \cite{feldmannTrackingExtendedObjects2011}. 
Approximating $\hat{h}_m(\cdot)$ by its first-order Taylor series expansion around the object centroid yields the Jacobian matrix $ H_m= {\partial \hat{h}_m}/{\partial x_\mathrm{K}}|_{x_\mathrm{K}=\hat{x}_\mathrm{K}}$,
and the likelihood is computed as
\begin{align}
\label{GGIW p_z_x}
    p(z|x) \approx \mathcal{N}(z;\hat{h}_m(x_\mathrm{K}), R_m(x_\mathrm{E})).
\end{align}
The covariance matrix accounts for both measurement noise and object's extent, given by
\begin{align}
\label{R_m}
    R_m(x_\mathrm{E})=\mathrm{diag}(\varrho \breve{H}_m x_\mathrm{E} \breve{H}_m^\mathrm{T} + \mathrm{diag}(\sigma^2_r,\sigma^2_\phi), \sigma^2_{\dot{r}}),
\end{align}
where $\breve{H}_m$ denotes the upper-left $2\times2$ sub-matrix of $H_m$.

When an object generates a set of measurements $W$ via propagation path $\mathfrak{P}_m$, a Poisson rate scaling factor $\gamma_m$ is introduced to model path-dependent attenuation with $\gamma_0=1$ for the direct path and $\gamma_m \in (0,1]$ for multi-bounce paths. The corresponding likelihood is
\begin{align}
    \tilde{g}(W|x) = \mathcal{PS}(|W|;\gamma_m x_\gamma )\prod_{z\in W} \mathcal{N}(z;\hat{h}_m(x_\mathrm{K}), R_m(x_\mathrm{E})). \notag
\end{align}
Given the predicted density $p_+(x)=\mathcal{GGIW}(x;\chi_+)$, the posterior density $p(x|W)$ is computed via Bayes' rule. The unnormalized posterior is
\begin{align}
    \tilde{p}(x|W) &=  \mathcal{G}(x_\gamma;\alpha_+, \beta_+) \mathcal{N}(x_\mathrm{K};\mu_+,P_+) \mathcal{IW}(x_\mathrm{E};\nu_+,V_+) \notag\\
    &\times \mathcal{PS}(|W|;\gamma_{m} x_\gamma ) \prod_{z\in W} \mathcal{N}(z;\hat{h}_m(x_\mathrm{K}),R_m(x_\mathrm{E})).\notag
\end{align}
Following the derivation in \cite{beardMultipleExtendedTarget2016a,granstromEstimationMaintenanceMeasurementa,granstromPhdFilterTracking2012} and \hyperref[GGIW Parameters Update]{Appendix}, the posterior $p(x|W)$ is approximated by $\mathcal{GGIW}(x;\chi_W)$ with updated parameters:
\begin{gather}
    \alpha_W = \alpha_++|W|,\quad 
    \beta_W = \beta_+ +\gamma_m, \notag\\ 
    \mu_W =\mu_+ +K(\bar z - h_m(\mu_+)),\quad 
    P_W = P_+-K\Lambda K^\mathrm{T}, \notag\\
    \nu_W = \nu_++|W|-1,\quad 
    V_W = V_+ + \breve{H}_m^{-1} \breve{D} (\breve{H}_m^{-1})^\mathrm{T}/\varrho. \notag
\end{gather}
Here, $\bar z$ is the centroid of $W$, $K = P_+ H_m^\mathrm{T}\Lambda^{-1}$ is the Kalman gain, $\Lambda = H_m P_+ H_m^\mathrm{T} + R_m(\bar{E})/|W|$ is the innovation covariance calculated using the expected extent $\bar{E} = V_+/(\nu_+ - 3)$, and $\breve{D}$ is the upper-left $2\times2$ submatrix of the scatter matrix $D=\sum_{z\in W} (z-\bar{z})(z-\bar{z})^\mathrm{T}$.

The marginal likelihood required for the GLMB update is
\begin{align}
\label{integral of GGIW}
    \int p_+(x)\tilde{g}(W|x) \mathrm{d}x = \eta_\gamma \eta_z \eta_R^{\frac{1-|W|}{2}}\eta_{\mathcal{IW}}
\end{align}
with components
\begin{gather}
    \eta_\gamma = \frac{\beta^\alpha \Gamma(\alpha+|W|) \gamma_m^{|W|}}{\Gamma(\alpha) (\beta+\gamma_m)^{(\alpha+|W|)}|W|!},
    \eta_z = \mathcal{N}(\bar{z};\hat{h}_m(\mu_+),\Lambda), \notag\\
    \eta_{R} = (\sigma_{r}^2 \sigma_{\phi}^2 \sigma_{\dot r}^2) \big|\bar{E}^{-1} + \varrho \breve{H}_m^\mathrm{T}\mathrm{diag}(\sigma_{r}^{-2},\sigma_{\phi}^{-2})\breve{H}_m \big|, \notag\\
    \eta_{\mathcal{IW}} = \frac{(2\pi)^{\frac{3(1-|W|)}{2}} 2^{|W|-1} |V_+|^{\frac{\nu_+}{2}} \Gamma_2(\frac{\nu_W}{2})}{|V_W|^{\frac{\nu_W}{2}} \Gamma_2(\frac{\nu_+}{2})} e^{\mathrm{tr}(D \Omega)/2}, \notag
\end{gather}
where $\Omega = \mathrm{diag}( \breve{E}^{-1} \mathrm{diag}(\sigma_{r}^{2},\sigma_{\phi}^{2}) \breve{E}^{-1},-\sigma_{\dot{r}}^{-2})$, and $\breve{E} = \varrho \breve{H}_m\bar{E}  \breve{H}_m^\mathrm{T}$. 
This formulation shows that when a prior GGIW density $p_+(x)$ is updated by multiple measurement sets as in \eqref{GLMB update}, and the object detection probabilities in \eqref{approximated psi} are independent of $x$, the resulting posterior $p_{\mathcal{U}(Z)}^{(\xi,\theta,\varphi)}(x,\ell)$ remains a GGIW density.

\subsection{Particle Update for Reflectors}

For reflectors modeled by the uniform stick representation, the gamma parameters $(\alpha,\beta)$ of the measurement rate $x_\gamma$ are updated using the closed-form expressions in Section~\hyperref[GGIW Update for Objects]{III-E}, whereas the kinematic and extent states require a different update strategy. 
Reflective points $\check{x}$ are assumed to be uniformly distributed along the line segment defined by the endpoints in \eqref{c_start c_end}. 
Accordingly, the position of a reflective point is expressed as $\check{x}_{c}^{\mathrm{p}}= x_\mathrm{K}^{\mathrm{p}} + c u_\mathrm{t}(\mathfrak{h})$, where $c$ is uniformly distributed on $[\mathfrak{s},\mathfrak{e}]$, and $u_\mathrm{t}(\mathfrak{h}) = \left[\cos(\mathfrak{h}),\sin(\mathfrak{h})\right]^\mathrm{T}$ is the unit tangent vector. 
The corresponding unit normal vector is $u_\mathrm{n}(\mathfrak{h}) = \left[-\sin(\mathfrak{h}),\cos(\mathfrak{h})\right]^\mathrm{T}$.
The likelihood \eqref{p_z_x} becomes
\begin{align}
\label{Stick p_z_x}
    p(z|x)
    &= \mathcal{N}(\dot{r};\hat{h}_m^{\dot{r}}(x_\mathrm{K}),\sigma^2_{\dot{r}}) \\
    &\quad \times \frac{1}{\mathfrak{e}-\mathfrak{s}}\int_{\mathfrak{s}}^{\mathfrak{e}} \mathcal{N}(z^{r\phi};\hat{h}_m^{r\phi}(\check{x}_{c}^{\mathrm{p}}),\mathrm{diag}(\sigma^2_r,\sigma^2_\phi)) \mathrm{d}c.\notag
\end{align}
Since this integral has no closed-form solution, and the elliptical Gaussian approximation in \eqref{GGIW p_z_x} is inaccurate over the reflector’s spatial extent, particle filtering is adopted to estimate the kinematic and extent states of reflectors. 

First, \eqref{Stick p_z_x} is approximated by a closed-form expression to enable efficient particle weight updates \cite{meyerScalableDetectionTracking2021,dingFusion2024}. 
The Gaussian integral is approximated in Cartesian coordinates as
\begin{align}
    &\int_{\mathfrak{s}}^{\mathfrak{e}} \mathcal{N}(\cdots) \mathrm{d}c \approx \int_{\mathfrak{s}}^{\mathfrak{e}} \mathcal{N}(z_\mathrm{xy};\check{x}_{c}^{\mathrm{p}},R_\mathrm{xy}) \mathrm{d}c, \notag
\end{align}
where the Cartesian position is obtained via the inverse measurement function $z_\mathrm{xy}=\hat{h}^{-1}_m(z^{r\phi})$, and the approximated covariance is $R_\mathrm{xy} = \breve{H}_m^{-1} \mathrm{diag}(\sigma^2_r,\sigma^2_\phi) (\breve{H}_m^{-1})^\mathrm{T}$. 
Assuming the reflector extent is significantly larger than the  measurement noise standard deviation, the multivariate Gaussian can be further approximated by two independent Gaussian components \cite{meyerScalableDetectionTracking2021}:
\begin{align}
    \int_{\mathfrak{s}}^{\mathfrak{e}} \mathcal{N}(z_\mathrm{xy};&\check{x}_{c}^{\mathrm{p}},R_\mathrm{xy}) \mathrm{d}c 
    \approx \mathcal{N}(d_z;0,\sigma_\mathrm{n}^2) \int_{\mathfrak{s}}^{\mathfrak{e}} \mathcal{N} (c; \mu_\mathrm{t}, \sigma^2_\mathrm{t})\mathrm{d}c \notag\\
    & = \mathcal{N}(d_z;0,\sigma_\mathrm{n}^2) \left[ \mathrm{Q}(\frac{\mathfrak{s}-\mu_\mathrm{t}}{\sigma_\mathrm{t}}) - \mathrm{Q}(\frac{\mathfrak{e}-\mu_\mathrm{t}}{\sigma_\mathrm{t}})\right],\notag
\end{align}
where $\mathrm{Q}(\cdot)$ denotes the Q-function \cite{borjessonQfunction1979}, and 
\begin{equation}
\begin{gathered}
\label{stick update variables}
    d_z = |u_\mathrm{n}(\mathfrak{h}) \cdot \Delta z |,\quad 
    \sigma_\mathrm{n}^2 = u_\mathrm{n}(\mathfrak{h})^\mathrm{T} R_\mathrm{xy} u_\mathrm{n}(\mathfrak{h}), \\
    \mu_\mathrm{t} = u_\mathrm{t}(\mathfrak{h}) \cdot \Delta z,\quad 
    \sigma_\mathrm{t}^2 = u_\mathrm{t}(\mathfrak{h})^\mathrm{T} R_\mathrm{xy} u_\mathrm{t}(\mathfrak{h}).
\end{gathered}
\end{equation}
Here, $\Delta z = h^{-1}_m(z^{r\phi})- x_\mathrm{K}^{\mathrm{p}}$, $d_z$ denotes the distance between $z_\mathrm{xy}$ and the reflector, $\mu_\mathrm{t}$ is the projection of $\Delta z$ on the reflector, and $\sigma_\mathrm{n}^2$ and $\sigma_\mathrm{t}^2$ are the projected noise variance along the normal and tangent directions, respectively \cite{meyerScalableDetectionTracking2021}. 
As a result, the likelihood \eqref{Stick p_z_x} yields a closed-form approximation.

Next, the prior kinematic and extent states are represented by a set of particles $\mathcal{X}_-=\{(x_\mathrm{K-}^{(l)},x_\mathrm{E-}^{(l)},\lambda_-^{(l)})\}_{l=1}^{L}$
with weights satisfying $\sum_n \lambda_-^{(l)}=1$. 
Predicted particles $\mathcal{X}_+$ are sampled from the transition densities \eqref{kinematic predict} and \eqref{particle transition density}, i.e.,
\begin{align}
    x_\mathrm{K+}^{(l)} \sim \mathcal{N}(F_+ x_\mathrm{K-}^{(l)},G_+Q_+G_+^\mathrm{T}),\
    x_\mathrm{E+}^{(l)} \sim \mathcal{N}(x_\mathrm{E-}^{(l)},Q_\mathrm{E+}), \notag
\end{align}
and  $\lambda_+^{(l)} = \lambda_-^{(l)}$.
When the reflector is detected via path $\mathfrak{P}_m$ with measurements $W$, particle weights are sequentially updated. 
Assume that the measurements are indexed in an arbitrary order as $W = \{z^{(\zeta)}\}_{\zeta=1}^{|W|}$. 
For the $\zeta$-th measurement, the weights are updated as $\lambda^{(l,\zeta)}= {\tilde\lambda_+^{(l,\zeta)}}/{\sum_{l=1}^L \tilde\lambda_+^{(l,\zeta)}}$, where
\begin{align}
    \tilde\lambda_{+}^{(l,\zeta)}=& \lambda_{+}^{(l,\zeta-1)} \mathcal{N}(\dot{r}^{(\zeta)};\hat{h}_m^{\dot{r}}(x_{\mathrm{K}+}^{(l)}),\sigma^2_{\dot{r}}) \notag\\
    &\times \frac{\mathcal{N}(d_z;0,\sigma_\mathrm{n}^2)}{\mathfrak{e}_+^{(l)}-\mathfrak{s}_+^{(l)}} \left[ \mathrm{Q}(\frac{\mathfrak{s}_+^{(l)}-\mu_\mathrm{t}}{\sigma_\mathrm{t}}) - \mathrm{Q}(\frac{\mathfrak{e}_+^{(l)}-\mu_\mathrm{t}}{\sigma_\mathrm{t}})\right]. \notag
\end{align}
Here, quantities $(d_z,\sigma_\mathrm{n}^2,\mu_\mathrm{t},\sigma_\mathrm{t}^2)$ are calculated by substituting $z^{(\zeta)}$, $x_\mathrm{K+}^{(l)}$, and $x_\mathrm{E+}^{(l)}$ into \eqref{stick update variables}. 
After processing all measurements, the updated particles are obtained as $\mathcal{X}=\{(x_\mathrm{K}^{(l,|W|)},x_\mathrm{E}^{(l,|W|)},\lambda^{(l,|W|)})\}_{l=1}^{L}$.
To mitigate weight degeneracy, systematic resampling with a Metropolis-Hastings step \cite{elfringParticleFiltersHandsOn2021} is applied to the updated particles when the number of effective particles falls below a threshold, i.e., $1/\sum_l(\lambda^{(l,|W|)})^2 < L_\mathrm{eff}$.
Finally, the marginal likelihood is approximated by $\prod_{\zeta=1}^{|W|} \sum_{l=1}^L \tilde\lambda_+^{(l,\zeta)}$ \cite{dingFusion2024,elfringParticleFiltersHandsOn2021}.

\section{Efficient Filter Implementation}
\label{Section IV}

This section details the implementation and pseudocode of the proposed MPET-GLMB filter.

\subsection{$\delta$-GLMB Joint Prediction and Update}
\label{Joint Prediction and Update}

The separated $\delta$-GLMB prediction and update recursion in Section~\hyperref[GLMB Filtering Recursion]{III-B} computes the prediction density $\boldsymbol{\pi}_+(\mathbf{X})$ independently of the measurements, which typically results in a large number of low-weight GLMB components and requires inefficient intermediate truncation \cite{voEfficientImplementationGeneralized2017}. 
To mitigate this issue, the joint prediction and update strategy in \cite{wangMultipledetectionMultitargetTracking2019} is adopted, in which target birth, death, and survival events are jointly enumerated together with measurement associations. 
By substituting the predicted weights in \eqref{w_+} into \eqref{w_Z}, the joint posterior density is obtained as
\begin{align}
\label{joint posterior}
    \boldsymbol{\pi}(\mathbf{X}|Z) \propto& \Delta(\mathbf{X}) \sum_{{\substack{\xi, I, I_+, S, \mathcal{U}, \theta,\varphi}}} w^{(I,\xi)} \\ 
    &\times  w_{\mathcal{U}}^{(I,I_+,\xi,\theta,\varphi)} \delta_{I_+}(\mathcal{L}(\mathbf{X})) [p_{\mathcal{U}}^{(\xi,\theta,\varphi)}]^\mathbf{X},\notag
\end{align}
where the joint weight accounts for target birth and survival probabilities with measurement likelihoods is
\begin{align}
\label{updated w}
    w_{\mathcal{U}(Z)}^{(I,I_+,\xi,\theta,\varphi)}=&[\mathit{1}_{\mathbb{B_+}} r_{\mathrm{B}+}]^{I_+\cap \mathbb{B}_+} [1-r_{\mathrm{B}+}]^{\mathbb{B}_+ - I_+} \\
    &\times [\bar{P}_\mathrm{S}^{(\xi)}]^{I_+\cap I} [1-\bar{P}_\mathrm{S}^{(\xi)}]^{I-I_+} [\eta_{\mathcal{U}(Z)}^{(\xi,\theta,\varphi)}]^{I_+},\notag
\end{align}
since \eqref{w_Z} can be rewritten as
\begin{align}
w_{\mathcal{U}}^{(I_+,\xi,\theta,\varphi)} &\propto w_+^{(I_+,\xi)} [\eta_{\mathcal{U}}^{(\xi,\theta,\varphi)}]^{I_+} \notag\\
&= \sum_{I\in\mathcal{F}(\mathbb{L})} w^{(I,\xi)} w_{\mathcal{U}}^{(I,I_+,\xi,\theta,\varphi)}. \notag
\end{align}
The posterior \eqref{joint posterior} indicates that each prior GLMB component $(I,\xi)$ generates a set of child components $(I,\xi,I_+,\theta,\varphi)$ during the joint prediction and update step. 
As the number of components grows exponentially, truncation is performed at each time step by retaining only those with the largest weights, which minimizes the $L_1$ approximation error \cite{voLabeledRandomFinite2013}. 

Given a GLMB component $(I,\xi)$ and a measurement partition $\mathcal{U}=\{\mathcal{U}_1,...,\mathcal{U}_{S}\}$, the labels are enumerated as
\begin{gather}
    I=I^\mathrm{o}\uplus I^{\mathrm{r}} =\{\ell_1:\ell_{N_I^\mathrm{o}}\}\uplus \{\ell_{N_I^\mathrm{o}+1}:\ell_{N_I}\},\notag\\
    \mathbb{B}_+ =\mathbb{B}_+^\mathrm{o}\uplus\mathbb{B}_+^\mathrm{r} =\{\ell_{N_I+1}:\ell_{N_B^\mathrm{o}}\}\uplus\{\ell_{N_B^\mathrm{o}+1}:\ell_{N}\}. \notag
\end{gather}
Here, $I^\mathrm{o}$ and $I^\mathrm{r}$ denote the label sets of existing objects and reflectors, respectively.
The label sets of newborn objects and reflectors are denoted by $\mathbb{B}_+^\mathrm{o}$ and $\mathbb{B}_+^\mathrm{r}$, respectively. 
The number of reflectors is given by $N^\mathrm{r}=N_I-N_I^\mathrm{o} + N-N_B^\mathrm{o}$.

For each $(I_+,\theta,\varphi)\in \mathcal{F}(\mathbb{L}\uplus\mathbb{B}_+)\times \Theta^{\mathcal{U}}_{\mathbf{X},\tilde{\mathbf{T}}}$, a unified association matrix $\vartheta$ of size $N\times (3N^\mathrm{r}+1)$ is introduced:
\begin{gather}
\begin{aligned}
\label{vartheta}
    \vartheta_{i,j}=
    \begin{cases}
        \theta(\ell_i) & \text{if } j=1, \ell_i\in I_+,\\
        \varphi((\ell_i,\ell_{j^\ast},m)) & \text{if } j>1, \ell_i\in I_+, \ell_i\neq\ell_{j^\ast},\\
        0 & \text{if } j>1, \ell_i\in I_+, \ell_i=\ell_{j^\ast},\\
        -1 & \text{otherwise},
    \end{cases}
\end{aligned}
\raisetag{13pt}
\end{gather}
where $m=\lceil(j-1)/N^\mathrm{r}\rceil$ ($\lceil\cdot\rceil$ is the ceiling function) identifies the propagation path, and $j^\ast=N_I^\mathrm{o}+j-(m-1) N^\mathrm{r}-1$ indexes the reflector. 
In this representation, $\vartheta_{i,j}=-1$ signifies target death or non-existence, $\vartheta_{i,j}=0$ indicates target existence without detection, and $\vartheta_{i,j}>0$ denotes an association between the target and a measurement subset. 

If a target $\ell_i$ exists ($\vartheta_{i,1}\geq 0$) but its multi-bounce path $(\ell_i,\ell_{j^\ast},m)$ can not be established, the corresponding entry in \eqref{vartheta} is still set to $\vartheta_{i,j}=0$, which is equivalent to an existing path with zero detection probability. 
If a target does not exist, none of its multi-bounce paths can exist. 
Consequently, the $i$-th row of $\vartheta$ satisfies $\vartheta_i\in \mathbb{E} \uplus \mathbb{N}$, where $\mathbb{E}= \{0:S\}^{3N^\mathrm{r}+1}$ represents target existence and $\mathbb{N}=\{-1\}^{3N^\mathrm{r}+1}$ represents target extinction.

With this definition, $\vartheta$ inherits the one-to-one (1-1) property of $\theta$ and $\varphi$, i.e., no distinct $(i,j)$ and $(i',j')$ with $\vartheta_{i,j}=\vartheta_{i',j'}>0$. 
Let $\mathcal{M}$ denote the set of all valid 1-1 mapping matrices $\vartheta$ with $\vartheta_i\in \mathbb{E} \uplus \mathbb{N}$. 
For any $\vartheta\in \mathcal{M}$, The corresponding $I_+$, $\theta$, and $\varphi$ are recovered as
\begin{equation}
\begin{gathered}
\label{I,theta,phi}
    I_+=\{\ell_i\in I\uplus\mathbb{B}_+: \vartheta_{i,1}\geq0\},\quad 
    \theta(\ell_i) = \vartheta_{i,1},\\
    \varphi(\ell_i,\ell_j,m) = \vartheta_{i,j-N_I^\mathrm{o}+1+(m-1)N^\mathrm{r}}.
\end{gathered}
\end{equation}
For $\mathfrak{r} \in \mathbb{E} \uplus \mathbb{N}$ and $\ell_i\in I\uplus\mathbb{B}_+$, the association weight is
\begin{gather}
\begin{aligned}
\label{association weight}
    \tilde{w}_i^{(I,\xi)}(\mathfrak{r}) = \begin{cases}
        1-\bar{P}_\xi(\ell_i) & \text{if } \mathfrak{r} \in \mathbb{N}, \\
        \bar{P}_\xi(\ell_i) \eta_{\mathcal{U}}^{(\xi,\mathfrak{r})}(\ell_i) & \text{if } \mathfrak{r} \in \mathbb{E}, \mathcal{V}_i(\mathfrak{r})=1, \\
        0 & \text{otherwise},
    \end{cases}
\end{aligned}
\raisetag{13pt}
\end{gather}
where $\bar{P}_\xi(\ell_i)$ is the survival probability $\bar{P}_\mathrm{S}^{(\xi)}$ for $\ell_i \in I$ or birth probability $r_{\mathrm{B}+}$ for $\ell_i \in \mathbb{B}_+$. 
The indicator $\mathcal{V}_i(\mathfrak{r})$ enforces valid associations by preventing self-reflection and duplicate measurement assignments:
\begin{align}
    \mathcal{V}_i(\mathfrak{r})=\begin{cases}
        0& \text{if } \exists i'\neq i,\mathfrak{r}^{(i)}=\mathfrak{r}^{(i')}>0; \text{or if } i>N_I^\mathrm{o},\\ &\quad \exists m\in M, \mathfrak{r}^{(i-N_I^\mathrm{o}+1+(m-1)N^\mathrm{r})}>0\\
        1& \text{otherwise}.
    \end{cases}\notag
\end{align}
The normalization term is
\begin{subequations}
\begin{align}
    \eta_{\mathcal{U}(Z)}^{(\xi,\mathfrak{r})}(\ell_i) &=\int \bar{p}_+^{(\xi)}(x,\ell) \hat\Psi_{\mathcal{U}(Z)}^{(\mathfrak{r})}(x,\ell) \mathrm{d}x,\\
    \label{psi_r}
    \hat\Psi_{\mathcal{U}(Z)}^{(\mathfrak{r})}(\mathbf{x}) &=
        \psi_{\mathrm{D},\mathcal{U}(Z)}^{(\mathfrak{r}^{(1)})}(\mathbf{x})
            \quad \prod_{\mathclap{\substack{j=2,...,\mathrm{dim}(\mathfrak{r}),\ m=\lceil(j-1)/N^\mathrm{r}\rceil\\ j^\ast=N_I^\mathrm{o}+j-(m-1) N^\mathrm{r}-1}}}\quad\ 
            \hat\psi_{\mathrm{R},\mathcal{U}(Z)}^{(\mathfrak{r}^{(j)},\ell_{j^\ast},m)}(\mathbf{x}),
\end{align}
\end{subequations}
where $\mathrm{dim}(\cdot)$ denotes vector dimension. 

\textit{Proposition 3}: For any $(I_+,\theta,\varphi)\in \mathcal{F}(\mathbb{L}\uplus\mathbb{B}_+)\times \Theta^{\mathcal{U}}_{\mathbf{X},\tilde{\mathbf{T}}}$ and its equivalent representation $\vartheta$, the joint component weight is the product of association weights
\begin{align}
\label{updated w in association weight}
    w_{\mathcal{U}}^{(I,I_+,\xi,\theta,\varphi)} = \mathit{1}_{\mathcal{M}}(\vartheta)\prod_{i=1}^{N} \tilde{w}_i^{(I,\xi)}(\vartheta_i).
\end{align}

\textit{Proof}: From \eqref{I,theta,phi}, comparison of \eqref{psi} and \eqref{psi_r} gives $\hat\Psi_{\mathcal{U}}^{(\vartheta_i)}=\hat\Psi_{\mathcal{U}}^{(\theta,\varphi)}$ and $\eta_{\mathcal{U}}^{(\xi,\vartheta_i)}=\eta_{\mathcal{U}}^{(\xi,\theta,\varphi)}$. 
Using \eqref{association weight} yields
\begin{align}
    \prod_{i=1}^{N_I} \tilde{w}_i^{(I,\xi)}(\vartheta_i)&= [1-\bar{P}_\mathrm{S}^{(\xi)}]^{I-I_+}
        [\bar{P}_\mathrm{S}^{(\xi)} \eta_{\mathcal{U}}^{(\xi,\theta,\varphi)}]^{I\cap I_+}, \notag\\
    \prod_{\mathclap{i=N_I+1}}^{N}\tilde{w}_i^{(I,\xi)}(\vartheta_i)&= [1-r_{\mathrm{B}+}]^{\mathbb{B}_+- I_+}
        [r_{\mathrm{B}+}\eta_{\mathcal{U}}^{(\xi,\theta,\varphi)}]^{\mathbb{B}_+\cap I_+}.\notag
\end{align}
Combining the two expressions recovers \eqref{updated w in association weight} from \eqref{updated w}, completing the proof. 

Proposition~3 shows that selecting high-weight child components $(I_+,\theta,\varphi)$ is equivalent to selecting 1-1 association matrices $\vartheta\in\mathcal{M}$ that maximize the product $\prod_{i=1}^{N} \tilde{w}_i^{(I,\xi)}(\vartheta_i)$.

\begin{algorithm}[t]
\caption{Gibbs Sampling}
\label{Alg1}

\KwIn{$\vartheta^{(1)}$, $T$, $\tilde{w}^{(I,\xi)}=[\tilde{w}_{i}^{(I,\xi)}(\mathfrak{r})]_{i=1:N,\mathfrak{r}\in\mathbb{E} \uplus \mathbb{N}}$}
\KwOut{$\vartheta^{(1)},...,\vartheta^{(T)}$}

\For{$t=2:T$}{
    \For{$i=1:N$}{
        \For{$\mathfrak{r}\in\mathbb{E} \uplus \mathbb{N}$}{
            \If{any positive entry of $\mathfrak{r}$ is in $\vartheta_{\bar{i}}^{(t)}$}{
                $\tilde{w}_i^{(I,\xi)}(\mathfrak{r})\gets 0$;
            }
        }
        $\vartheta_i^{(t)}\sim \mathrm{Categorical}(\mathbb{E} \uplus \mathbb{N},\tilde{w}_i^{(I,\xi)})$;
    }
    $\vartheta^{(t)}\gets[\vartheta^{(t)}_1,...,\vartheta^{(t)}_N]$.
}

\end{algorithm}

\subsection{Gibbs Sampling}

Ranked assignment methods such as Murty's algorithm~\cite{millerMurty1997} can be employed to extract the $T$ most significant child components of each GLMB component. 
However, these methods become computationally prohibitive in high-dimensional settings, and the ranked order is unnecessary for the MPET-GLMB filter. 
Instead, the association mapping $\vartheta$ is treated as a realization of a random variable with distribution $\pi(\vartheta)\propto \mathit{1}_{\mathcal{M}}(\vartheta)\prod_{i=1}^{N} \tilde{w}_i^{(I,\xi)}(\vartheta_i)$ as proposed in \cite{voEfficientImplementationGeneralized2017}.
This formulation assigns nonzero probabilities only to valid mappings and favors mappings with higher weights.

A block Gibbs sampler is used to sample $\vartheta$ row by row, with a Markov transition kernel
\begin{align}
    \pi(\vartheta'|\vartheta)
        &= \prod_{n=1}^{N} \pi_n(\vartheta'_n|\vartheta'_{1:n-1},\vartheta_{n+1,N})\notag\\
        &= \prod_{n=1}^{N} \frac{\pi(\vartheta'_{1:n-1},\vartheta_{n+1,N})}{\sum_{\vartheta_n}\pi(\vartheta'_{1:n-1},\vartheta_n,\vartheta_{n+1,N})}.
\end{align}
For each $n\in\{1,\dots,N\}$, define $\vartheta_{\bar{n}}=(\vartheta_{1:n-1},\vartheta_{n+1:N})$. 
Following Proposition~3 in \cite{voEfficientImplementationGeneralized2017} and Proposition~6 in \cite{nguyenTrackingCellsTheir2021}, the conditional distribution is
\begin{align}
    \pi_n(\vartheta_{n}|\vartheta_{\bar{n}})\propto \begin{cases}
        \tilde{w}_n^{(I,\xi)}(\vartheta_n)& \text{if no } i\in\vartheta_{n}, j\in\vartheta_{\bar{n}},\\ &\quad \text{satisfy }i=j>0, \\
        0& \text{otherwise}.
    \end{cases}
\end{align}
The sampling procedure is summarized in Algorithm~\ref{Alg1}. 
In terms of computational complexity, Murty's algorithm requires $O(T([S+1)^{3N^\mathrm{r}+1} + N]^3)$ operations \cite{millerMurty1997}, whereas the proposed Gibbs sampler reduces the complexity to $O(TN^2 [(S+1)^{3N^\mathrm{r}+1}+1])$, since each row samples a categorical distribution with cost $O(N[(S+1)^{3N^\mathrm{r}+1}+1])$.

\subsection{MPET-GLMB Filter}
\subsubsection{Joint Prediction and Update Implementation}

Following \cite{voEfficientImplementationGeneralized2017}, the prior density \eqref{prior GLMB} is rewritten as
\begin{align}
\label{prior GLMB with h}
    \boldsymbol{\pi}(\mathbf{X}) = \Delta(\mathbf{X})\sum_{h=1}^H w^{(h)} \delta_{I^{(h)}}(\mathcal{L}(\mathbf{X})) [p^{(h)}]^\mathbf{X},
\end{align}
where $w^{(h)}= w^{(I^{(h)},\xi^{(h)})}$, $p^{(h)}= p^{(\xi^{(h)})}$.
The original GLMB parameters $w^{(I,\xi)}$, $p^{(\xi)}$, and $(I,\xi)\in \mathcal{F}(\mathbb{L})\times\Xi$ are enumerated as $\{(I^{(h)},\xi^{(h)},w^{(h)},p^{(h)})\}_{h=1}^H$. 
After replacing the component index $(I,\xi)$ by $h$, the predicted single-target density and survival probability are
\begin{gather}
    \bar{p}_+^{(h)}(x,\ell_i) = \bar{p}_+^{(\xi^{(h)})}(x,\ell_i),\quad
    \bar{P}_\mathrm{S}^{(h)}(\ell_i) = \bar{P}_\mathrm{S}^{(\xi^{(h)})}(\ell_i). \notag
\end{gather}

Since the number of measurement partitions grows combinatorially with the size of the measurement set \cite{beardMultipleExtendedTarget2016a}, exhaustive enumeration in \eqref{joint posterior} is generally intractable. 
Thus, clustering methods (e.g., distance-based clustering \cite{granstromPhdFilterTracking2012} and DBSCAN \cite{khanDBSCAN2014}) are applied to obtain a set of likely partitions, denoted by $\mathcal{P}(Z)=\{\mathcal{U}^{(u)}\}_{u=1}^{|\mathcal{P}(Z)|}$. 
The association weight in \eqref{association weight} becomes
\begin{gather}
\begin{aligned}
\label{w_h}
    \tilde{w}_i^{(u,h)}(\mathfrak{r}) =\begin{cases}
        1-\bar{P}_h(\ell_i) & \text{if } \mathfrak{r} \in \mathbb{N}^{(h)}, \\
        \bar{P}_h(\ell_i) \eta_{Z}^{(u,h,\mathfrak{r})}(\ell_i) & \text{if } \mathfrak{r} \in \mathbb{E}^{(u,h)}, \mathcal{V}_i(\mathfrak{r})=1 \\
        0 & \text{otherwise},
    \end{cases}
\end{aligned}
\raisetag{13pt}
\end{gather}
where $\bar{P}_h(\ell_i)$ is the survival probability $\bar{P}_\mathrm{S}^{(h)}$ for $\ell_i \in I$ or birth probability $r_{\mathrm{B}+}$ for $\ell_i \in \mathbb{B}_+$, $\eta_{Z}^{(u,h,\mathfrak{r})}(\ell_i) = \eta_{\mathcal{U}^{(u)}}^{(\xi^{(h)},\mathfrak{r})}(\ell_i)$, $\mathbb{N}^{(h)}=\{-1\}^{3|\mathcal{R}(I^{(h)})|+1}$, $\mathbb{E}^{(u,h)}=\{0:|\mathcal{U}^{(u)}|\}^{3|\mathcal{R}(I^{(h)})|+1}$, with $\mathcal{R}(I)$ denoting the set of reflector labels in $I$.

Algorithm~\ref{Alg2} summarizes the computation of posterior $\delta$-GLMB parameters $\{(I_Z^{(h_Z)},w_Z^{(h_Z)},p_Z^{(h_Z)})\}_{h_Z=1}^{H_Z}$, where $\xi^{(h)}$ is omitted for brevity as it is implicit in $h$. 
From \eqref{GLMB update}, \eqref{I,theta,phi}, and \eqref{updated w in association weight}, each updated component indexed by $(u,h,t)$ has 
\begin{gather}
    I_Z^{(u,h,t)} = \{\ell_i\in I^{(h)}\uplus\mathbb{B}_+: \vartheta_{i,1}^{(u,h,t)}\geq0\}, \notag\\
    \label{I_ht} w_Z^{(u,h,t)} \propto w^{(h)} \prod_{i=1}^{|I^{(h)}\uplus \mathbb{B}_+|} \tilde{w}_i^{(u,h)}(\vartheta_i^{(u,h,t)}),\\
    p_Z^{(u,h,t)}(x,\ell_i) = {\bar{p}_+^{(h)}(x,\ell_i) \hat\Psi_{\mathcal{U}^{(u)}}^{(\vartheta_i^{(h,t)})}(x,\ell_i)}/{\eta_{Z}^{(u,h,\vartheta_i^{(h,t)})}(\ell_i)}.\notag
\end{gather}
Duplicated GLMB components are merged by summing their weights, after which the remaining weights are normalized to obtain the posterior density.

\begin{algorithm}[t]
\caption{Joint Prediction and Update}
\label{Alg2}

\KwIn{$\{(I^{(h)},w^{(h)},p^{(h)})\}_{h=1}^H$, $Z$, $H_\mathrm{max}$}
\vspace{2pt}
\KwOut{$\{(I_Z^{(h_Z)},w_Z^{(h_Z)},p_Z^{(h_Z)})\}_{h_Z=1}^{H_Z}$}

Sample $[T^{(h)}]_{h=1}^H$ from a multinomial distribution with $H_\mathrm{max}$ trails and event probabilities of $[w^{(h)}]_{h=1}^H$;

Generate partitions $\mathcal{P}(Z)=\{\mathcal{U}^{(u)}\}$ via clustering;

\For{$u=1: |\mathcal{P}(Z)|$, $h=1:H$}{
    Initialize $\vartheta^{(u,h,1)}$;
    
    Compute weights $\tilde{w}^{(u,h)}=\{\tilde{w}_{i}^{(u,h)}(\mathfrak{r})\}$ using \eqref{w_h};

    $\{\vartheta^{(u,h,t)}\}_{t=1}^{\tilde{T}^{(u,h)}} \gets \mathrm{Unique}(\mathrm{Gibbs}(\vartheta^{(u,h,1)},T^{(h)},\tilde{w}^{(u,h)}))$;\footnotemark

    \For{$t=1:\tilde{T}^{(u,h)}$}{
        Compute $I_Z^{(u,h,t)}$, $w_Z^{(u,h,t)}$, $p_Z^{(u,h,t)}$ using \eqref{I_ht};
    }
}

$(\{(I_Z^{(h_Z)},p_Z^{(h_Z)})\}_{h_Z=1}^{H_Z},[U_{u,h,t}]) \gets \mathrm{Unique}(\{(I_Z^{(u,h,t)},p_Z^{(u,h,t)})\}_{(u,h,t)=(1,1,1)}^{(|\mathcal{P}(Z)|,H,\tilde{T}^{(u,h)})})$;

\For{$h_Z=1:H_Z$}{
    $w_Z^{(h_Z)} \gets \sum_{u,h,t: U_{u,h,t}=h_Z}w_Z^{(u,h,t)}$;
}

Normalize weights $\{w_Z^{(h_Z)}\}_{h_Z=1}^{H_Z}$.

\end{algorithm}

\footnotetext{
$\{\cdot\}$ represents a MATLAB cell data structure which may contain duplicated elements; $\mathrm{Unique}(\cdot)$ returns the unique elements of a cell, and also the indices of these elements in the original cell if required.}

\subsubsection{Measurement Driven Adaptive Birth Model}

Existing multipath RFS-based trackers \cite{yangMultipathGeneralizedLabeled2018a,liuRFSBasedMultipleExtended2023,wangMultipledetectionMultitargetTracking2019} rely on hand-crafted target birth density and fixed multipath propagation models, which limit their applicability in real-world traffic environments. 
In contrast, the proposed MPET-GLMB filter jointly estimates object and reflector states, enabling online refinement of the multipath model. Moreover, a measurement-driven birth model is incorporated to enhance adaptability in dynamic scenarios. 
At each scan, the measurement set is partitioned into stationary and moving subsets, $Z=Z_\mathrm{s} \uplus Z_\mathrm{m}$, where the stationary measurements $Z_\mathrm{s}$ satisfy a Doppler threshold $|\dot{r}|<\dot{r}_\mathrm{min}$. 
All measurements are transformed into Cartesian coordinates, denoted by $Z^\mathrm{C}=Z_\mathrm{s}^\mathrm{C} \uplus Z_\mathrm{m}^\mathrm{C}$. 
Hough transform \cite{ILLINGWORTH198887} is then applied to $Z_\mathrm{s}^\mathrm{C}$ to detect lines. 
Stationary measurements within distance $d_\mathrm{max}$ of any detected line are retained as potential reflector measurements $Z_\mathrm{r}^\mathrm{C}$. 
Finally, $Z_\mathrm{r}^\mathrm{C}\uplus Z_\mathrm{m}^\mathrm{C}$ is clustered to generate feasible measurement partitions.

After computing the posterior $\delta$-GLMB density, newborn targets are generated adaptively from measurements rather than predefined birth models. 
Each measurement subset in every feasible partition is assigned a unique new label, forming the birth label set $\mathbb{B}_+$ that satisfies $\mathbb{B}_+\cap \mathbb{L}=\emptyset$. 
For $W_\mathrm{B} \subseteq Z_\mathrm{r}\uplus Z_\mathrm{m}$, the associated birth label is denoted $\ell_{W_\mathrm{B}}$. 
The likelihood that $W_\mathrm{B}$ initializes a new target is defined as
\begin{align}
    \tilde{r}_\mathrm{B}(\ell_{W_\mathrm{B}}) = 1 - \frac{\sum_{u,h,t} \mathit{1}_{\mathcal{U}^{(u,h,t)}_Z } (W_\mathrm{B}) w_Z^{(u,h,t)}}{\sum_{u,h,t}  w_Z^{(u,h,t)}}, \notag
\end{align}
where $\mathcal{U}^{(u,h,t)}_Z$ denotes the measurement subset associated with $\vartheta^{(u,h,t)}$, and $\mathit{1}_{\mathcal{U}^{(u,h,t)}_Z } (W_\mathrm{B})$ indicates whether $W_\mathrm{B}$ is assigned to an existing target \cite{deuschLabeledMultiBernoulliFilter2014,linMeasurementDrivenBirth2016}. 
If $W_\mathrm{B}$ is used in all hypotheses, then $\tilde{r}_\mathrm{B}(W_\mathrm{B})=0$, meaning no new target can be initialized. 
The resulting birth density \eqref{birth LMB} has existence probabilities
\begin{align}
   r_{\mathrm{B}+}(\ell_{W_\mathrm{B}})= \begin{cases}
       r_{\mathrm{B_{max}}} \tilde{r}_\mathrm{B}(\ell_{W_\mathrm{B}}) & \text{if } \tilde{r}_\mathrm{B}(\ell_{W_\mathrm{B}})>r_\mathrm{B_{min}},\\
       0 & \text{if } \tilde{r}_\mathrm{B}(\ell_{W_\mathrm{B}})\leq r_\mathrm{B_{min}},
   \end{cases}\notag
\end{align}
where $r_{\mathrm{B_{max}}}\in[0,1]$ is the maximum existence probability, and $r_{\mathrm{B_{min}}}$ is the initialize threshold.

The spatial extent of a newborn target initialized from $W_\mathrm{B}$ is estimated from the measurement distribution. 
Let $\bar{W}_\mathrm{B}^\mathrm{C}$ and $C_{W_\mathrm{B}}$ denote the mean and covariance of the measurement locations in Cartesian coordinates, respectively. 
Following the Gaussian approximation of elliptical extents \cite{feldmannTrackingExtendedObjects2011}, the extent matrix is estimated as $C_{W_\mathrm{B}} / \varrho$. 
Eigenvalue decomposition of the extent yields the heading vector $u_{W_\mathrm{B}}$ associated with the largest eigenvalue $d_{W_\mathrm{B}}$.

If $W_\mathrm{B}$ is stationary and aligned with a detected line, a newborn reflector is initialized using a particle set $\mathcal{X}_{W_\mathrm{B}}=\{(x_{\mathrm{K},W_\mathrm{B}}^{(l)},x_{\mathrm{E},W_\mathrm{B}}^{(l)},\lambda_{W_\mathrm{B}}^{(l)})\}_{l=1}^{L}$, where
\begin{gather}
    \lambda_{W_\mathrm{B}}^{(l)}=1/L,\quad x_{\mathrm{K},W_\mathrm{B}}^{(l)} \sim \mathcal{N}(\bar{W}_\mathrm{B}^\mathrm{C},G_+ Q_+ G_+^\mathrm{T}),\notag\\
    x_{\mathrm{E},W_\mathrm{B}}^{(l)} \sim \mathcal{N}([\angle(u_{W_\mathrm{B}}),-d_{W_\mathrm{B}}/2, d_{W_\mathrm{B}}/2]^\mathrm{T},Q_\mathrm{E+}).\notag
\end{gather}
Otherwise, a new object is initialized with a GGIW density
\begin{align}
    p_{\mathrm{B}+}(x,\ell_{W_\mathrm{B}}) &= \mathcal{GGIW}(x;\chi_{W_\mathrm{B}}) = \mathcal{G}(x_\gamma;\alpha_{W_\mathrm{B}},\beta_{W_\mathrm{B}})\notag\\
        & \times \mathcal{N}(x_\mathrm{K};\mu_{W_\mathrm{B}},P_{\mathrm{B}}) \mathcal{IW}(x_\mathrm{E};\nu_{\mathrm{B}},V_{W_\mathrm{B}}), \notag
\end{align}
where the kinematic covariance $P_{\mathrm{B}}$ and inverse Wishart degrees of freedom $\nu_{\mathrm{B}}$ are predefined, and
\begin{gather}
    \alpha_{W_\mathrm{B}} = |W_\mathrm{B}|^2/\gamma_\mathrm{var},\quad 
    \beta_{W_\mathrm{B}} = |W_\mathrm{B}|/\gamma_\mathrm{var},\notag\\
    \mu_{W_\mathrm{B}} = \bar{W}_\mathrm{B}^\mathrm{C},\quad 
    V_{W_\mathrm{B}} = C_{W_\mathrm{B}}/[\varrho(\nu_{\mathrm{B}}+3)],\notag
\end{gather}
with $\gamma_\mathrm{var}$ denoting the measurement rate variance.

\subsubsection{Track Extraction and Pruning}

After each joint prediction and update step, labeled target estimates are extracted from the posterior $\delta$-GLMB density.
The cardinality is first estimated using the \textit{maximum a posteriori} criterion, and the highest-weight GLMB component with this cardinality is selected, denoted by $(I_Z^{(h_\mathrm{max})},p_Z^{(h_\mathrm{max})})$.
For each target labeled by $\ell\in I_Z^{(h_\mathrm{max})}$, its kinematic and extent states are estimated from the distribution $p_Z^{(h_\mathrm{max})}(\cdot,\ell)$.
To suppress unlikely hypotheses, the posterior $\delta$-GLMB is pruned by discarding all components with weights below the threshold $w_Z^{(h)} < w_\mathrm{min}$.
Further implementation details can be found in \cite{beardMultipleExtendedTarget2016a,voOverviewMultiObjectEstimation2024}.

\section{Experiments and Performance Analysis}
\label{Section V}

This section evaluates the performance of the proposed MPET-GLMB filter in two simulated traffic scenarios and one real-world scenario.
The method is benchmarked against the state-of-the-art RFS-based multipath multiple extended target tracker (MP-ET-PHD)~\cite{liuRFSBasedMultipleExtended2023} and a conventional GLMB filter.
Since the GLMB and MP-ET-PHD filters rely on fixed object-birth models, they are initialized with newborn components near the true object-birth positions.
For ablation studies, a variant of MPET-GLMB using the same fixed birth model is also evaluated, which is referred to as MPET-GLMB-FB. 
All filters use the GGIW extended object model in Section~\ref{Section III} and share identical parameters unless otherwise specified.

Multi-target state estimation accuracy is evaluated using the optimal sub-pattern assignment (OSPA) distance \cite{beardOSPA2Using2017}, which captures both localization and cardinality errors between the estimated state set $\hat{X}$ and the ground-truth set $X$.
Since OSPA is designed for unlabeled sets and does not account for target identities, the OSPA$^{(2)}$ distance is additionally employed to evaluate trajectory-level errors \cite{beardOSPA2Using2017}.
Moreover, the widely used CLEAR MOT metrics \cite{bernardinEvaluatingMultipleObject2008, luitenHOTA2021}, including multi-object tracking accuracy (MOTA), multi-object tracking precision (MOTP), true positives (TP), false positives (FP), false negatives (FN), and ID switches (IDS), are reported in the following evaluation.
For the OSPA and OSPA$^{(2)}$, the cutoff distance is set to $2\,\mathrm{m}$ with an order of 1, and OSPA$^{(2)}$ window length is set to 5. 
For CLEAR MOT evaluation, a distance threshold of $1\,\mathrm{m}$ is used to determine true positive matches.

\subsection{Scenario 1}

\begin{figure}[t!]
    \centering
    \includegraphics[width=0.95\linewidth]{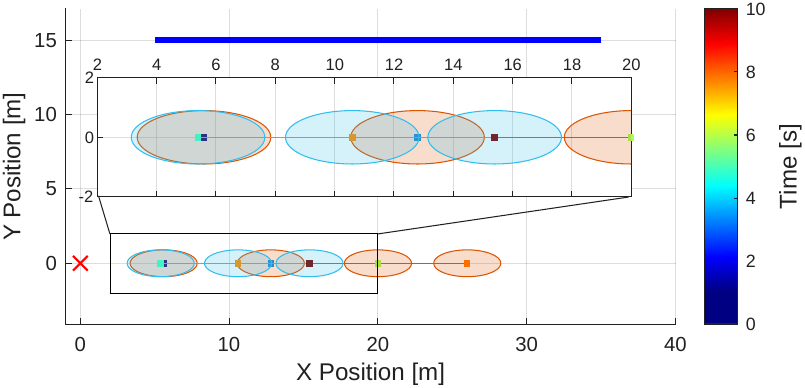}
    \caption{Simulated ground truth for Scenario 1. Red cross denotes radar position. Blue line represents the reflector. Square markers show object centroids at selected time steps, with ellipsoids denoting object extents. Objects are assigned different colors for identification.}
    \label{scene1_truth}
\end{figure}

Scenario 1 simulates the occlusion between objects to evaluate the filter's ability to exploit multipath measurements for robust tracking.
As shown in Fig.~\ref{scene1_truth}, a stationary reflector of length $30\,\mathrm{m}$ is positioned at $[20\,\mathrm{m},15\,\mathrm{m}]^\mathrm{T}$. 
Two objects with dimensions $4.5\,\mathrm{m}\times1.8\,\mathrm{m}$ move away from the radar under a constant velocity model. The initial kinematic states are $[5\,\mathrm{m},0\,\mathrm{m},3\,\mathrm{m/s},0\,\mathrm{m/s}]^\mathrm{T}$ and $[5\,\mathrm{m},0\,\mathrm{m},2\,\mathrm{m/s},0\,\mathrm{m/s}]^\mathrm{T}$, respectively.
The first object enters the observation area at $1\,\mathrm{s}$ and exits at $8\mathrm{s}$.
The second object enters at $5\,\mathrm{s}$ and remains until the simulation ends at $10\,\mathrm{s}$.
When the second object appears, the first object becomes occluded, causing its detection probabilities via the direct and type 1 two-bounce paths to drop by $90\%$. 

The radar operates at a scan rate of $5\,\mathrm{Hz}$ with measurement noise standard deviations $\sigma_r=0.2\,\mathrm{m}$, $\sigma_\phi=0.5^\circ$, and $\sigma_{\dot r}=0.5\,\mathrm{m/s}$.
Clutter is uniformly distributed over  $(r,\phi)\in[0,60\,\mathrm{m}]\times[-90^\circ,90^\circ]$ with a Poisson rate of $\gamma_c=20$.
For unoccluded objects, the direct path detection probability is $P_\mathrm{d}=0.95$, while the multi-bounce path detection probability is $P_{\mathrm{d,R}}=0.9$.
The measurement rates $x_\gamma$ of the reflector and two objects are set to $80$, $30$, and $30$, respectively, with a Poisson rate scaling factor of $\gamma_{\mathrm{m}\in\{1,2,3\}}=0.6$.

The birth parameters of the MPET-GLMB filter are configured as $r_\mathrm{B_{max}}=1$, $r_\mathrm{B_{min}}=0.5$, $\gamma_\mathrm{var}=25$, $\nu_\mathrm{B}=10$, and $P_\mathrm{B}=\mathrm{diag}(1,1,4,4,0.03)$.
In contrast, the other three filters use a single newborn component with an existence probability of $0.02$.
Its GGIW parameters are $\chi_\mathrm{B}=(\alpha_\mathrm{B},\beta_\mathrm{B},\mu_\mathrm{B},P_\mathrm{B},\nu_\mathrm{B},V_\mathrm{B})$, where $\alpha_\mathrm{B}=\bar x_\gamma^2/\gamma_\mathrm{var}$, $\beta_\mathrm{B}=\bar x_\gamma/\gamma_\mathrm{var}$, $\mu_\mathrm{B}=[4,0,0,0]^\mathrm{T}$, $V_\mathrm{B}= 4\bar E/(\nu_\mathrm{B}+3)$.
Here, $\bar x_\gamma=40$ and $\bar{E}=\mathrm{diag}(4,2)$ are the expected measurement rate and extent matrix for the object, respectively.
The object survival probability is set to $P_\mathrm{S}=0.99$. 
The MP-ET-PHD filter is provided with accurate prior environmental knowledge by setting its multipath measurement model with the exact reflector position and shape.
All filters employ the same measurement-clustering procedure, and unused stationary measurements are removed for the GLMB and MP-ET-PHD filters.

\begin{figure}[t!]
    \centering
    \includegraphics[width=1\linewidth]{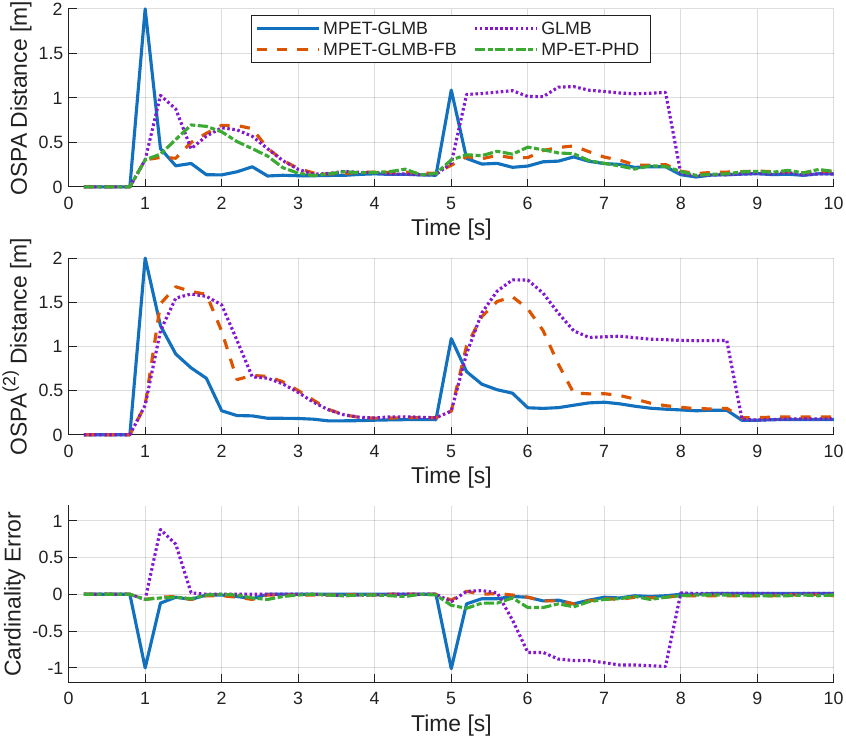}
    \caption{OSPA, OSPA$^{(2)}$, and cardinality error averaged over 100 Monte Carlo runs for Scenario 1.}
    \label{scene1_ospa}
\end{figure}

Tracking performance is evaluated over 100 Monte Carlo runs. 
As shown in Table~\ref{table scene 1} and Fig.~\ref{scene1_ospa}, MPET-GLMB achieves the highest overall tracking performance among all methods.
Although the measurement-driven adaptive birth model requires at least one frame before initializing a new track, causing transient cardinality estimation errors and OSPA peaks when newborn objects appear at $1\,\mathrm{s}$ and $5\,\mathrm{s}$, MPET-GLMB quickly converges to accurate and stable track estimations.
Consequently, it attains the lowest average OSPA and OSPA$^{(2)}$ distances and fewest ID-switch errors.
In contrast, the MPET-GLMB-FB and MP-ET-PHD filters achieve more accurate cardinality estimation, but their fixed birth components compete with existing tracking hypotheses, resulting in significantly higher ID-switch rates. 
The GLMB filter performs worst due to the absence of a multipath measurement model, leading to the highest cardinality error and OSPA distances. 
Specifically, GLMB fails to maintain the track of the first object following its occlusion, whereas the other three filters utilize multipath measurements to obtain reliable estimations.
Finally, although MP-ET-PHD achieves OSPA and MOTP performance comparable to MPET-GLMB-FB, it does not estimate target IDs or trajectories. 
Consequently, the OSPA$^{(2)}$, MOTA, and IDS metrics relying on object IDs are not reported for this filter.

\begin{table}[t!]
\centering
\belowrulesep=0pt
\aboverulesep=0pt
\caption{Tracking Performance Metrics (Scenario 1)}
\label{table scene 1}
\begin{threeparttable}
\renewcommand\tabcolsep{0.9pt}
\begin{tabular}{c|cccccccc}
\toprule
    \textbf{Method} & \textbf{OSPA} & \textbf{OSPA}$^{(2)}$ & \textbf{MOTA} & \textbf{MOTP} & \textbf{TP} & \textbf{FP} & \textbf{FN} & \textbf{IDS} \\
\midrule
    \textbf{MPET-GLMB}       & \textbf{0.2291} & \textbf{0.3437} & \textbf{0.9444} & \textbf{0.1677} & 5777 & \textbf{11} & 323 & \textbf{5} \\
    MPET-GLMB-FB  & 0.2657 & 0.5519 & 0.8844 & 0.2427 & \textbf{5922} & 42 & \textbf{178} & 484 \\
    GLMB            & 0.4814 & 0.7418 & 0.6307 & 0.2371 & 4817 & 500 & 1283 & 470 \\
    MP-ET-PHD       & 0.2588 & / & / & 0.2191 & 5855 & \textbf{11} & 245 & / \\
\bottomrule
\end{tabular}
\begin{tablenotes}
    \footnotesize
    \item[*] TP, FP, FN, and IDS are cumulative sums over 100 Monte Carlo runs. Other metrics are average values. OSPA, OSPA$^{(2)}$, and MOTP are measured in meters. \textbf{Bold} denotes the best result.
\end{tablenotes}
\end{threeparttable}
\end{table}

\subsection{Scenario 2}
\label{Scenario 2}

\begin{figure}[t]
    \centering
    \includegraphics[width=0.95\linewidth]{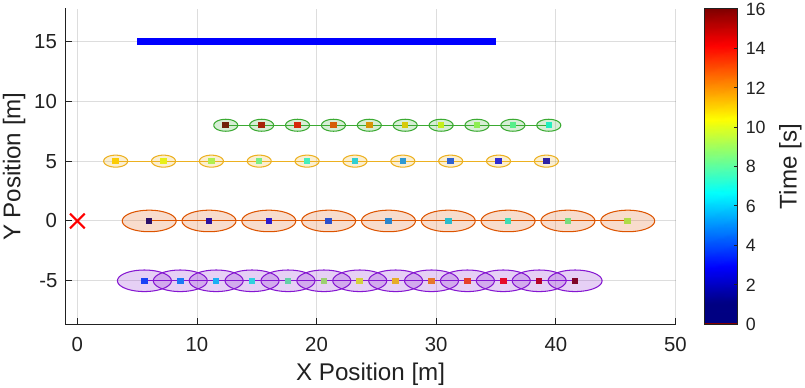}
    \caption{Simulated ground truth for Scenario 2.}
    \label{scene2_truth}
\end{figure}

\begin{table}[t]
\centering
\belowrulesep=0pt
\aboverulesep=0pt
\caption{Object Simulation Parameters (Scenario 2)}
\label{table scene 2 truth}
\begin{threeparttable}
\renewcommand\tabcolsep{1.5pt}
\begin{tabular}{c|ccccc}
\toprule
    \textbf{ID} & \textbf{Initial State} & \textbf{Dimensions} & $x_\gamma$ & \textbf{Birth} & \textbf{Death} \\
\midrule
    1   & $[5\,\mathrm{m},0\,\mathrm{m},5\,\mathrm{m/s},0\,\mathrm{m/s}]^\mathrm{T}$ & $4.5\,\mathrm{m}\times1.8\,\mathrm{m}$ & 30 & $1\,\mathrm{s}$ & $9\,\mathrm{s}$ \\
    2   & $[5\,\mathrm{m},-5\,\mathrm{m},3\,\mathrm{m/s},0\,\mathrm{m/s}]^\mathrm{T}$ & $4.5\,\mathrm{m}\times1.8\,\mathrm{m}$ & 30 & $4\,\mathrm{s}$ & $16\,\mathrm{s}$ \\
    3   & $[40\,\mathrm{m},5\,\mathrm{m},-4\,\mathrm{m/s},0\,\mathrm{m/s}]^\mathrm{T}$ & $2\,\mathrm{m}\times1\,\mathrm{m}$ & 20 & $2\,\mathrm{s}$ & $11\,\mathrm{s}$ \\
    4   & $[40\,\mathrm{m},8\,\mathrm{m},-3\,\mathrm{m/s},0\,\mathrm{m/s}]^\mathrm{T}$ & $2\,\mathrm{m}\times1\,\mathrm{m}$ & 20 & $7\,\mathrm{s}$ & $16\,\mathrm{s}$ \\
\bottomrule
\end{tabular}
\end{threeparttable}
\end{table}

\begin{figure}[t!]
    \centering
    \includegraphics[width=0.9\linewidth]{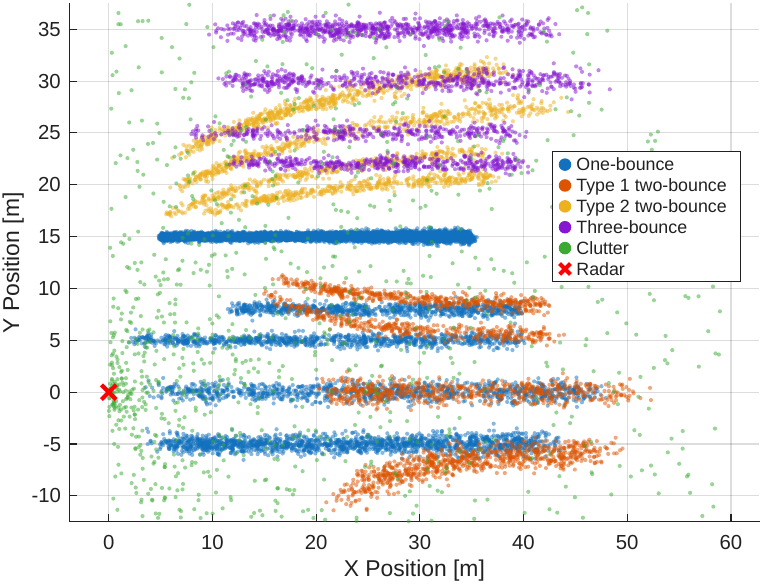}
    \caption{All measurements in one Monte Carlo run for Scenario 2.}
    \label{scene2_meas}
\end{figure}

\begin{table}[t]
\centering
\belowrulesep=0pt
\aboverulesep=0pt
\caption{Tracking Performance Metrics (Scenario 2)}
\label{table scene 2}
\begin{threeparttable}
\renewcommand\tabcolsep{0.4pt}
\begin{tabular}{c|cccccccc}
\toprule
    \textbf{Method} & \textbf{OSPA} & \textbf{OSPA}$^{(2)}$ & \textbf{MOTA} & \textbf{MOTP} & \textbf{TP} & \textbf{FP} & \textbf{FN} & \textbf{IDS} \\
\midrule
    \textbf{MPET-GLMB}       & \textbf{0.2505} & \textbf{0.3804} & \textbf{0.8920} & \textbf{0.1632} & \textbf{18194} & \textbf{872} & \textbf{1206} & \textbf{17} \\
    MPET-GLMB-FB  & 0.3885 & 0.6122 & 0.6956 & 0.1913 & 17540 & 3511 & 1860 & 534 \\
    GLMB            & 0.6908 & 0.8686 & 0.3995 & 0.1943 & 17391 & 9129 & 2009 & 512 \\
    MP-ET-PHD       & 0.4664 & / & / & 0.1826 & 15541 & 1145 & 3859 & / \\
\bottomrule
\end{tabular}
\begin{tablenotes}
    \footnotesize
    \item[*] TP, FP, FN, and IDS are cumulative sums over 100 Monte Carlo runs. Other metrics are average values. OSPA, OSPA$^{(2)}$, and MOTP are measured in meters. \textbf{Bold} denotes the best result.
\end{tablenotes}
\end{threeparttable}
\end{table}

\begin{figure}[t]
    \centering
    \includegraphics[width=1\linewidth]{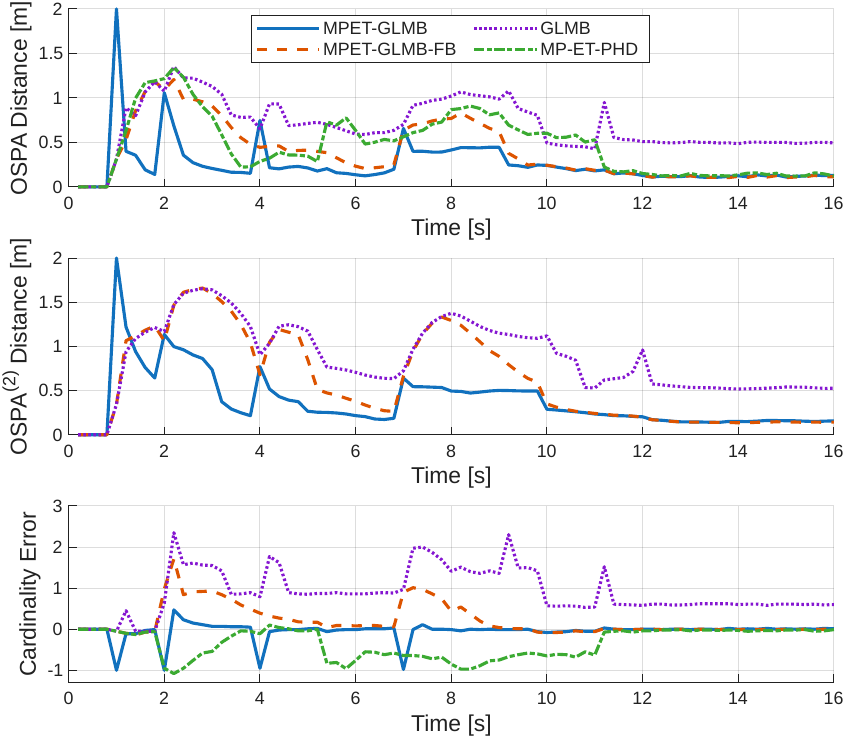}
    \caption{OSPA, OSPA$^{(2)}$, and cardinality error averaged over 100 Monte Carlo runs for Scenario 2.}
    \label{scene2_ospa}
\end{figure}

Scenario 2 considers a more complex environment in which four objects of distinct shapes enter the observation area from different positions, as detailed in Fig.~\ref{scene2_truth} and Table~\ref{table scene 2 truth}. 
As illustrated in Fig.~\ref{scene2_meas}, the type-1 two-bounce multipath measurements overlap with the direct path measurements, posing significant challenges to the filters. 
Object occlusion is not simulated in this scenario. 
The fixed birth model comprises four components, each initialized at the true object birth location with an existence probability of $0.01$.
For the two smaller objects, the expected extent matrix and measurement rate are set to $\bar E=\mathrm{diag}(2,1)$ and $\bar x_\gamma=30$, respectively. 
All other parameters are identical to those used in Scenario 1.

As shown in Table~\ref{table scene 2} and Fig.~\ref{scene2_ospa}, the proposed MPET-GLMB filter achieves superior tracking performance across all metrics in this challenging scenario. 
Despite incorporating multipath measurement models, MPET-GLMB-FB and MP-ET-PHD exhibit large cardinality errors, underscoring the effectiveness of the proposed adaptive birth model in complex multipath environments.
GLMB and MPET-GLMB-FB tend to overestimate cardinality because type-1 two-bounce measurements repeatedly trigger births near fixed birth components.
However, MPET-GLMB-FB mitigates false tracks using its multipath model, resulting in lower OSPA and cardinality errors than GLMB. 
In comparison, the MP-ET-PHD filter underestimates cardinality due to the Poisson assumptions and inherent limitations of PHD-based approximation of the multi-target density \cite{lundquistGGIWCPHD2013}. 
The consistently low OSPA, MOTP, and IDS demonstrate that MPET-GLMB  maintains stable and accurate trajectory estimates.

\subsection{Scenario 3}

\begin{figure}[t]
    \centering
    \includegraphics[width=1\linewidth]{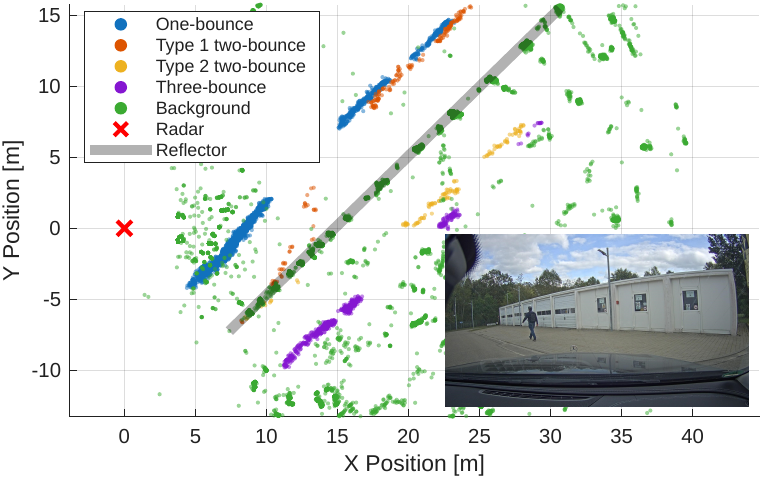}
    \caption{All radar measurements for Scenario 3. The experimental environment is shown in the lower right corner of the image.}
    \label{scene3_meas}
\end{figure}

\begin{table}[t]
\centering
\belowrulesep=0pt
\aboverulesep=0pt
\caption{Tracking Performance Metrics (Scenario 3)}
\label{table scene 3}
\begin{threeparttable}
\renewcommand\tabcolsep{1.4pt}
\begin{tabular}{c|cccccccc}
\toprule
    \textbf{Method} & \textbf{OSPA} & \textbf{OSPA}$^{(2)}$ & \textbf{MOTA} & \textbf{MOTP} & \textbf{TP} & \textbf{FP} & \textbf{FN} & \textbf{IDS} \\
\midrule
    \textbf{MPET-GLM}B       & \textbf{0.2147} & \textbf{0.3076} & \textbf{0.9375} & \textbf{0.0957} & 105 & \textbf{0} & \textbf{7} & \textbf{0} \\
    MPET-GLMB-FB  & 0.2153 & 0.4218 & 0.8482 & 0.1011 & 109 & 7 & 3 & 7 \\
    GLMB            & 0.2273 & 0.5140 & 0.8214 & 0.1045 & \textbf{110} & 10 & 2 & 8 \\
    MP-ET-PHD       & 0.4301 & / & / & 0.1093 & 93 & \textbf{0} & 19 & / \\
\bottomrule
\end{tabular}
\begin{tablenotes}
    \footnotesize
    \item[*] OSPA, OSPA$^{(2)}$, and MOTP are measured in meters. \textbf{Bold} denotes the best result.
\end{tablenotes}
\end{threeparttable}
\end{table}

\begin{figure}[t!]
    \centering
    \includegraphics[width=1\linewidth]{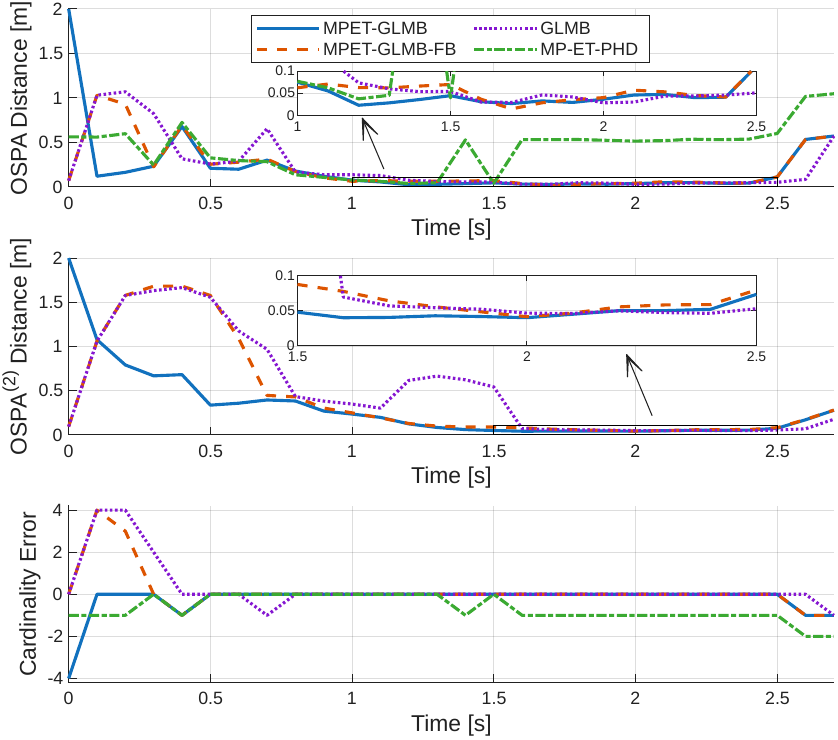}
    \caption{OSPA, OSPA$^{(2)}$, and cardinality error for Scenario 3.}
    \label{scene3_ospa}
\end{figure}

Radar measurements for Scenario 3 are sourced from the Radar Ghost dataset~\cite{krausRadarGhostDataset2021}, which focuses on automotive radar ghost targets and provides detailed annotations of multipath measurements, as illustrated in Fig.~\ref{scene3_meas}. 
The selected sequence\footnote{File name: \raggedright{scenario-13\_sequences-01-05-05-06\_start-frames-013-123-166-019\_cycl-ped-ped-ped\_train.h5}} contains three pedestrians and one cyclist moving between an automotive radar and a reflective wall. 
The radar scan rate is $10\,\mathrm{Hz}$, and the mean Cartesian position of each object's one-bounce measurements is used as the ground truth at each scan.

To accommodate the higher variability of real-world radar measurements, the measurement rate variance is increased to $\gamma_\mathrm{var}=64$, and the inverse Wishart degrees of freedom are reduced to $\nu_\mathrm{B}=4$. 
The fixed birth model comprises four components, each with an existence probability of $0.01$, and is initialized at the true birth positions. 
The expected extent matrix and measurement rate are set to $\bar E=\mathrm{diag}(0.5,0.5)$ and $\bar x_\gamma=20$, respectively. 
Since all objects appear at the beginning of the sequence, birth events are restricted to the first three frames for all filters to reduce ID-switch errors. 
The multipath detection probability is set to $P_{\mathrm{d,R}}=0.8$, while all other parameters follow those in Scenarios 1 and 2.

As shown in Fig.~\ref{scene3_ospa} and Table~\ref{table scene 3}, MPET-GLMB outperforms the other filters in both estimation accuracy and track ID consistency within this real-world scenario.  
Despite restricting births to the first three frames, filters with fixed birth models still exhibit cardinality overestimation and ID switches during the initialization, further demonstrating the advantage of the proposed adaptive birth model.  
Furthermore, while the three GLMB-based filters maintain accurate cardinality after the initial period, the MP-ET-PHD consistently underestimates the number of objects after $1.5\,\mathrm{s}$, which may pose safety concerns in real-world traffic applications.

\subsection{Ablation Study}

To assess the individual contributions of the uniform stick reflector model, Gibbs sampling-based truncation, and the adaptive birth model, ablation experiments are conducted in Scenario 2 with variants of the MPET-GLMB filter.  
As summarized in Table~\ref{table ablation}, MPET-GLMB-FB replaces the adaptive birth model with a fixed birth model (detailed in Section~\hyperref[Scenario 2]{V-B}), MPET-GLMB-GGIW replaces the uniform stick reflector model with a standard GGIW model, and MPET-GLMB-MA replaces the Gibbs sampling with Murty's algorithm. 

Reflector extent estimation accuracy is evaluated using the Hausdorff distance.  
Given a reflector extent state $x_\mathrm{E}$, let $\mathcal{C}=\{\mathrm{C}_\mathrm{start}, \mathrm{C}_\mathrm{end}\}$ denote the endpoints computed via \eqref{c_start c_end}.  
For MPET-GLMB-GGIW, the reflector heading and length are extracted from the eigenvalue decomposition of $x_\mathrm{E}$. 
The Hausdorff distance is given by
\begin{align}
    d_\mathrm{H}(\hat{\mathcal{C}},\mathcal{C}) = \max\left[\max_{\hat{\mathrm{C}}\in\hat{\mathcal{C}}} d_{\mathcal{C}}(\hat{\mathrm{C}}), \max_{{\mathrm{C}}\in\mathcal{C}} d_{\hat{\mathcal{C}}}(\mathrm{C})\right], \notag
\end{align} 
where $d_{\mathcal{C}}(\hat{\mathrm{C}})$ denotes the minimum distance from an estimated endpoint $\hat{\mathrm{C}}$ to the ground truth line segment.  
Computational efficiency is assessed using the average frames processed per second (FPS) on the same simulation platform.

As shown in Table~\ref{table ablation}, MPET-GLMB-MA achieves the highest tracking accuracy, slightly surpassing the proposed MPET-GLMB, as Murty's algorithm guarantees optimal assignments.   
However, its high computational cost substantially reduces the FPS, whereas the Gibbs sampling-based implementation provides an approximately 40 times speedup. 
Consistent with the analysis in Section~\hyperref[Scenario 2]{V-B}, MPET-GLMB-FB exhibits degraded accuracy due to the fixed birth model, with additional false tracks further reducing efficiency.   
Regarding reflector modeling, MPET-GLMB-GGIW produces significantly larger Hausdorff distances, indicating that the GGIW model inadequately captures the linear geometry of reflectors. 
These extent estimation errors also lead to increased ID switches. 
Although MPET-GLMB-GGIW achieves slightly higher FPS due to lower computational cost, the resulting accuracy loss is substantial.  
Overall, the ablation results confirm that each proposed module contributes as intended, and MPET-GLMB achieves an effective balance between tracking accuracy and real-time efficiency.

\begin{table}[t]
\centering
\belowrulesep=0pt
\aboverulesep=0pt
\caption{Tracking Performance Metrics (Ablation Study)}
\label{table ablation}
\begin{threeparttable}
\renewcommand\tabcolsep{1.5pt}
\begin{tabular}{c|cccccc}
\toprule
    \textbf{Method} & \textbf{OSPA}$\downarrow$ & \textbf{OSPA}$^{(2)}$$\downarrow$ & \textbf{MOTA}$\uparrow$ & \textbf{IDS}$\downarrow$ & $d_\mathrm{H}$$\downarrow$ & \textbf{FPS}$\uparrow$ \\
\midrule
    \textbf{MPET-GLMB}       & 0.2505 & 0.3804 & 0.8920 & 17 & 2.17 &  3.51\\
    MPET-GLMB-FB  & 0.3885 & 0.6122 & 0.6956 & 534 & 2.20 &  2.21\\
    MPET-GLMB-GGIW  & 0.2972 & 0.4831 & 0.8153 & 187 & 11.32 &  4.87 \\
    MPET-GLMB-MA  & 0.2466 & 0.3792 & 0.8969 & 16 & 2.13 & 0.089 \\
\bottomrule
\end{tabular}
\begin{tablenotes}
    \footnotesize
    \item[*] IDS is a cumulative sum over 100 Monte Carlo runs. Other metrics are average values. OSPA, OSPA$^{(2)}$, and $d_\mathrm{H}$ are measured in meters.
\end{tablenotes}
\end{threeparttable}
\end{table}

\section{Conclusion}
\label{Section VI}

This paper presented the multipath extended target generalized labeled multi-Bernoulli (MPET-GLMB) filter, a unified Bayesian framework for tracking multiple extended targets in dynamic multipath environments. 
By integrating the labeled RFS formulation with extended target modeling, the proposed method jointly resolves the uncertainties in target existence, measurement partitioning, and measurement-to-path association. 
A Gibbs sampling-based joint prediction and update strategy was developed to maintain computational tractability, together with a measurement-driven adaptive birth model to enable robust track initialization. 
Extensive experiments on simulated traffic scenarios and real-world automotive radar data demonstrated that the MPET-GLMB filter consistently outperforms state-of-the-art methods in state estimation accuracy and trajectory continuity, particularly under severe occlusion and dense target birth conditions.
Future work will investigate extensions to multi-sensor fusion and real-time deployment on embedded hardware platforms.

\section*{APPENDIX}
\label{GGIW Parameters Update}
Given the prior GGIW density $p(x)=\mathcal{GGIW}(x;\chi)$ and measurements $W$, the posterior density is $p(x|W) = \frac{\tilde{p}(x|W)}{\int \tilde{p}(x|W) \mathrm{d}x}$,
where the unnormalized density is
\begin{align}
    \tilde{p}(x|&W) = \ \mathcal{G}(x_\gamma;\alpha, \beta) \mathcal{N}(x_\mathrm{K};\mu,P) \mathcal{IW}(x_\mathrm{E};\nu,V) \notag\\
        &\times \mathcal{PS}(|W|;\gamma_{m} x_\gamma ) \prod_{z\in W} \mathcal{N}(z;\hat{h}_m(x_\mathrm{K}),R_m(x_\mathrm{E})). \notag
\end{align}
The product of gamma and Poisson PDFs is written as
\begin{align}
    \mathcal{G}(x_\gamma;&\alpha, \beta) \mathcal{PS}(|W|;\gamma_m x_\gamma ) \notag\\
        =& \frac{\beta^\alpha}{\Gamma(\alpha)}x_\gamma^{\alpha-1}e^{-\beta x_\gamma} \frac{(\gamma_m x_\gamma)^{|W|}e^{-\gamma_m x_\gamma}}{|W|!}  \notag\\
        =& \mathcal{G}(x_\gamma;\alpha_W,\beta_W) \eta_\gamma, \notag
\end{align}
where $\eta_\gamma =\frac{\beta^\alpha \Gamma(\alpha+|W|) \gamma_m^{|W|}}{\Gamma(\alpha) (\beta+\gamma_m)^{(\alpha+|W|)}|W|!}$, $\alpha_W =\alpha+|W|$, and $\beta_W = \beta+\gamma_m$.
Since $x_\gamma$ is independent of $x_\mathrm{K}$ and $x_\mathrm{E}$, the posterior density of the measurement rate becomes
\begin{align}
    p(x_\gamma|W)&= \frac{\eta_\gamma\ \mathcal{G}(x_\gamma;\alpha_W,\beta_W) }{\int \eta_\gamma\ \mathcal{G}(x_\gamma;\alpha_W,\beta_W) \mathrm{d}x_\gamma}=\mathcal{G}(x_\gamma;\alpha_W,\beta_W).\notag
\end{align}
It remains to compute $\tilde{p}(x_\mathrm{K},x_\mathrm{E}|W)=\mathcal{N}(x_\mathrm{K};\mu,P)\times \mathcal{IW}(x_\mathrm{E};\nu,V) \eta_{\mathcal{N}}(W,R_m) \prod_{z\in W}\mathcal{N}(z;\hat{h}_m(x_\mathrm{K}), R_m)$ and its integral $\int \tilde{p}(x_\mathrm{K},x_\mathrm{E}|W) \mathrm{d}x$.

Following \cite[Appendix A]{granstromPhdFilterTracking2012}, the product of Gaussian PDFs over $z\in W$ can be written as
\begin{align}
    \prod_{z\in W}&\mathcal{N}(z;...) = \eta_{\mathcal{N}}(W,R_m) \mathcal{N}(\bar z; \hat{h}_m(x_\mathrm{K}),\frac{R_m}{|W|}),\notag
\end{align}
where $\eta_{\mathcal{N}}(W,R_m) = (2\pi)^{\frac{3(1-|W|)}{2}} |R_m|^{\frac{1-|W|}{2}} e^{\mathrm{tr}(-\frac{1}{2}D R_m^{-1})}$ and $D = \sum_{z\in W} (z-\bar z)(z-\bar z)^\mathrm{T}$. Using the EKF approximation yields $\mathcal{N}(x_\mathrm{K};\mu,P) \mathcal{N}(\bar z; \hat{h}_m(x_\mathrm{K}),\frac{R_m}{|W|}) \approx \mathcal{N}(x_\mathrm{K};\mu_W,P_W)\mathcal{N}(\bar z; \hat{h}_m(\mu),\Lambda)$, where 
\begin{equation}
\begin{gathered}
\label{GGIW Kalman}
    \mu_W = \mu + K(\bar{z}-\hat{h}_m(\mu)),\ 
    P_W = P - K \Lambda K^\mathrm{T},\\
    K = P H_m^\mathrm{T}\Lambda^{-1},\ 
    \Lambda = H_m P H_m^\mathrm{T} + {R_m(\bar{E})}/{|W|}.
\end{gathered}
\end{equation}
Notably, the EKF innovation covariance $\tilde{\Lambda}=H_m P H_m^\mathrm{T} + {R_m(x_\mathrm{E})}/{|W|}$ originally depends on $x_\mathrm{E}$. Since ${R_m(x_\mathrm{E})}$ has minor effects on $\tilde{\Lambda}$ when $|W|$ is large, we approximate it by substituting $x_\mathrm{E}$ with the expectation $\bar{E}=V/(\nu-3)$ in \eqref{GGIW Kalman}. As $x_\mathrm{E}$ is symmetric positive-definite, the matrix determinant lemma yields the approximation:
\begin{align}
\label{approx R_m}
    |R_m|=&\sigma^2_{\dot{r}} |\varrho \breve{H}_m x_\mathrm{E} \breve{H}_m^\mathrm{T}+ \mathrm{diag}(\sigma^2_r,\sigma^2_\phi)| \notag\\
        =& \sigma^2_{\dot{r}}\sigma^2_r\sigma^2_\phi\  |x_\mathrm{E}^{-1} + \varrho\breve{H}_m^\mathrm{T} \mathrm{diag}(\sigma^{-2}_r,\sigma^{-2}_\phi)\breve{H}_m| \ |x_\mathrm{E}| \notag\\
        \approx& \sigma^2_{\dot{r}}\sigma^2_r\sigma^2_\phi\  |\bar{E}^{-1} + \varrho\breve{H}_m^\mathrm{T} \mathrm{diag}(\sigma^{-2}_r,\sigma^{-2}_\phi)\breve{H}_m| \ |x_\mathrm{E}|\notag\\
        =& \eta_R |x_\mathrm{E}|.\notag
\end{align}
We approximate $R_m^{-1}$ using the first-order Neumann series
\begin{align}
    &R_m^{-1}=\mathrm{diag}([\varrho \breve{H}_m x_\mathrm{E} \breve{H}_m^\mathrm{T} + \mathrm{diag}(\sigma^2_r,\sigma^2_\phi)]^{-1}, \sigma_{\dot{r}}^{-2} ) \notag\\
        &\approx \mathrm{diag}( (\varrho \breve{H}_m x_\mathrm{E} \breve{H}_m^\mathrm{T})^{-1} - \breve{E}^{-1} \mathrm{diag}(\sigma_{r}^{2},\sigma_{\phi}^{2}) \breve{E}^{-1}, \sigma_{\dot{r}}^{-2} ),\notag
\end{align}
where $\breve{E} = \varrho \breve{H}_m\bar{E}  \breve{H}_m^\mathrm{T}$.
Furthermore, we have
\begin{align}
    &\eta_{\mathcal{N}}(W,R_m) \mathcal{IW}(x_\mathrm{E};\nu,V)\notag\\
    &= (2\pi)^{\frac{3(1-|W|)}{2}} |R_m|^{\frac{1-|W|}{2}}  \frac{|V|^\frac{\nu}{2}|x_\mathrm{E}|^{-\frac{\nu+3}{2}} }{2^\nu \Gamma_2(\frac{\nu}{2})} e^{\frac{-\mathrm{tr}(D R_m^{-1} + V x_\mathrm{E}^{-1})}{2}}\notag\\
    &\approx  (2\pi)^{\frac{3(1-|W|)}{2}} \eta_R^{\frac{1-|W|}{2}}  \frac{|V|^\frac{\nu}{2}|x_\mathrm{E}|^{\frac{-(\nu+|W|+2)}{2}}}{2^\nu \Gamma_2(\frac{\nu}{2})} e^{\frac{\mathrm{tr}(D \Omega)}{2}} \notag\\
        &\quad \times e^{\frac{-\mathrm{tr}[(V+ \breve{H}_m^{-1}\breve{D}(\breve{H}_m^{-1})^\mathrm{T}/\varrho) x_\mathrm{E}^{-1}]}{2}}\notag\\
    &= \eta_R^{\frac{1-|W|}{2}}\ \eta_\mathcal{IW} \ \mathcal{IW}(x_\mathrm{E}; \nu_W, V_W),
\end{align}
where $\breve{D}$ is the upper-left $2\times2$ submatrix of $D$, $\Omega = \mathrm{diag}( \breve{E}^{-1} \mathrm{diag}(\sigma_{r}^{2},\sigma_{\phi}^{2}) \breve{E}^{-1},-\sigma_{\dot{r}}^{-2})$, $\nu_W = \nu +|W|-1$, $V_W = V+ \breve{H}_m^{-1}\breve{D}(\breve{H}_m^{-1})^\mathrm{T}/\varrho$, and $\mathrm{tr}(D R_m^{-1})$ is evaluated using the properties of matrix trace:
\begin{align}
    &\mathrm{tr}[ D \mathrm{diag}( (\varrho \breve{H}_m x_\mathrm{E} \breve{H}_m^\mathrm{T})^{-1} - \breve{E}^{-1} \mathrm{diag}(\sigma_{r}^{2},\sigma_{\phi}^{2}) \breve{E}^{-1}, \sigma_{\dot{r}}^{-2} ) ] \notag\\
     &= \mathrm{tr}[ (\breve{H}_m^{-1} \breve{D} (\breve{H}_m^{-1})^\mathrm{T}/ \varrho) x_\mathrm{E}^{-1}   ] - \mathrm{tr}(D \Omega).\notag
\end{align}
Consequently, $\mathcal{GGIW}(x|\chi_W)$ is the approximated posterior density, and the marginal likelihood is $\int \tilde{p}(x|W) \approx \eta_\gamma\mathcal{N}(\bar z; \hat{h}_m(\mu),\Lambda) \eta_R^{{(1-|W|)}/{2}} \eta_\mathcal{IW}$.

\bibliographystyle{IEEEtran}
\bibliography{bibtex/bib/IEEEabrv,bibtex/bib/IEEEexample}

\begin{IEEEbiography}[{\includegraphics[width=1in,clip,keepaspectratio]{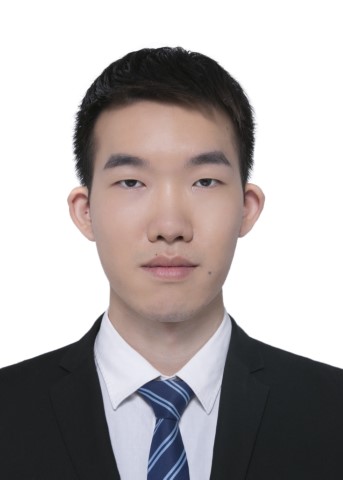}}]{Guanhua Ding}
(Student Member, IEEE) received the B.S. and M.Sc. degrees from Beihang University (BUAA), Beijing, China, in 2020 and 2023, respectively. He is currently pursuing the Ph.D. degree in signal and information processing with the School of Electronic Information Engineering, BUAA. His research interests include multi-object tracking, extended object tracking, multi-sensor data fusion, and random finite set theory.
\end{IEEEbiography}

\begin{IEEEbiography}[{\includegraphics[width=1in,clip,keepaspectratio]{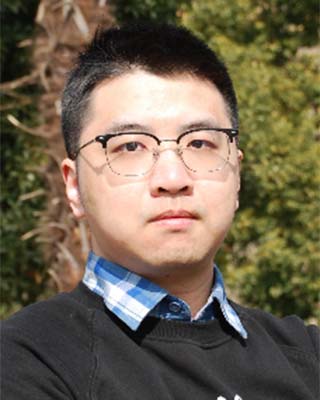}}]{Qinchen Wu}
(Student Member, IEEE) received the B.S. and M.Sc. degrees from Beihang University (BUAA), Beijing, China, in 2019 and 2021, respectively. He is currently working toward the Ph.D. degree in signal and information processing with the School of Electronic Information Engineering, Beihang University, Beijing, China. His research interests include random finite set, group target tracking, and multi-sensor data fusion.
\end{IEEEbiography}

\begin{IEEEbiography}[{\includegraphics[width=1in,clip,keepaspectratio]{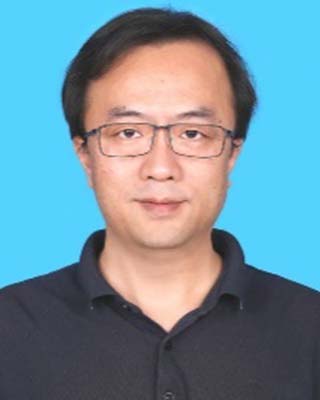}}]{Jinping Sun}
(Member, IEEE) received the M.Sc. and Ph.D. degrees from Beihang University (BUAA), Beijing, China, in 1998 and 2001, respectively. Currently, he is a Professor with the School of Electronic Information Engineering, BUAA. His research interests include statistical signal processing, high-resolution radar signal processing, target tracking, image understanding, and robust beamforming.
\end{IEEEbiography}

\begin{IEEEbiography}[{\includegraphics[width=1in,clip,keepaspectratio]{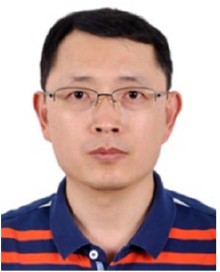}}]{Yanping Wang}
(Member, IEEE) received the B.S. and M.S. degrees in mechanical electronics engineering from the Beijing Institute of Technology, Beijing, China, in 1998 and 2001, respectively, and the Ph.D. degree in signal and information processing from the Institute of Electronics, Chinese Academy of Sciences, Beijing, in 2004. He is currently a Professor with the North China University of Technology, Beijing. His research interests include information fusion, ground-based radar imaging and deformation monitoring.
\end{IEEEbiography}

\begin{IEEEbiography}[{\includegraphics[width=1in,clip,keepaspectratio]{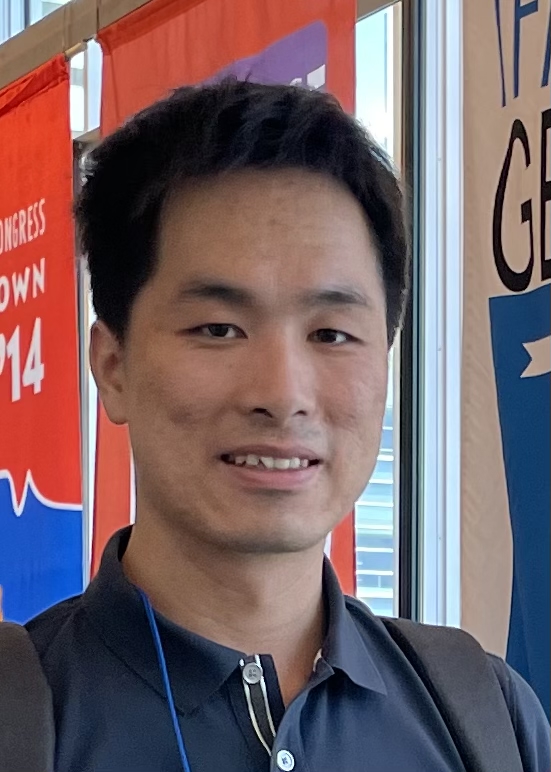}}]{Bing Zhu}
(Senior Member, IEEE) received the B.S. and Ph.D. degrees in control theory and applications from Beihang University, Beijing, China, in 2007 and 2013, respectively. He was with University of Pretoria, Pretoria, South Africa, as a Postdoctoral Fellow supported by Vice-Chancellor Postdoctoral Fellowship from 2013 to 2015. He was with Nanyang Technological University, Singapore, as a Research Fellow from 2015 to 2016. He joined Beihang University, as an Associate Professor in 2016. His research interests include model predictive control, smart sensing, and demand-side management for new energy systems. Dr Zhu serves as an Associate Editor for Acta Automatica Sinica.
\end{IEEEbiography}

\begin{IEEEbiography}[{\includegraphics[width=1in,clip,keepaspectratio]{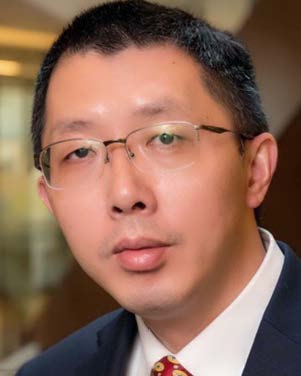}}]{Guoqiang Mao}
(Fellow, IEEE) is a Chair Professor and the Director of the Center for Smart Driving and Intelligent Transportation Systems, Southeast University. From 2014 to 2019, he was a Leading Professor, the Founding Director of the Research Institute of Smart Transportation, and the Vice-Director of the ISN State Key Laboratory, Xidian University. Before that, he was with the University of Technology Sydney and The University of Sydney. He has published 300 papers in international conferences and journals that have been cited more than 15,000 times. His H-index is 57 and was in the list of Top 2\% most-cited scientists worldwide by Stanford University in 2022, 2023, and 2024 both by Single Year and by Career Impact. His research interests include intelligent transport systems, the Internet of Things, wireless localization techniques, mobile communication systems, and applied graph theory and its applications in telecommunications. He is a fellow of AAIA and IET. He received the ``Top Editor" Award for outstanding contributions to IEEE Transactions on Vehicular Technology in 2011, 2014, and 2015. He has served as the chair, the co-chair, and a TPC member in several international conferences. He has been serving as the Vice-Director of Smart Transportation Information Engineering Society and Chinese Institute of Electronics since 2022. He was the Co-Chair of the IEEE ITS Technical Committee on Communication Networks from 2014 to 2017. He is an Editor of IEEE Transactions on Intelligent Transportation Systems (since 2018), IEEE Transactions on Wireless Communications (2014–2019), and IEEE Transactions on Vehicular Technology (2010–2020).
\end{IEEEbiography}

\end{document}